\documentclass[aps]{revtex4}
\usepackage{amsfonts}
\usepackage{amsmath}
\usepackage{amssymb,epsf}
\usepackage{graphicx,epsfig}
\usepackage{color}
\usepackage{graphicx}
\usepackage{epstopdf}

\begin{document}

\title{Geometrical thermodynamics and P-V criticality of the black holes\\
with power-law Maxwell field}
\author{S. H. Hendi$^{1,2}$\footnote{%
email address: hendi@shirazu.ac.ir}, B. Eslam Panah$^{1,2}$\footnote{%
email address: behzad.eslampanah@gmail.com}, S.
Panahiyan$^{1,3}$\footnote{ email address: sh.panahiyan@gmail.com}
and M. S. Talezadeh$^{1}$} \affiliation{$^1$ Physics Department
and Biruni Observatory, College of Sciences, Shiraz
University, Shiraz 71454, Iran\\
$^2$ Research Institute for Astronomy and Astrophysics of Maragha (RIAAM),
P.O. Box 55134-441 Maragha, Iran\\
$^3$Physics Department, Shahid Beheshti University, Tehran 19839, Iran}

\begin{abstract}
We study thermodynamical structure of Einstein black holes in the
presence of power Maxwell invariant nonlinear electrodynamics for
two different cases. The behavior of temperature and conditions
regarding to the stability of these black holes are investigated.
Since the language of geometry is an effective method in general
relativity, we concentrate on the geometrical thermodynamics to
build a phase space for studying thermodynamical properties of
these black holes. In addition, taking into account the
denominator of heat capacity, we use the proportionality between
cosmological constant and thermodynamical pressure to extract the
critical values for these black holes. Besides, the effects of
variation of different parameters on the thermodynamical structure
of these black holes are investigated. Furthermore, some
thermodynamical properties such as volume expansion coefficient,
speed of sound and isothermal compressibility coefficient are
calculated and some remarks regarding these quantities are given.
\end{abstract}

\maketitle

\section{Introduction}

In the early 20th century, it was found that some principal
questions could not be resolved by Newton's theory. One of the
most notable problems was the prediction of Mercury orbit's
orientation. According to these deficits, in 1915 Einstein
published a tensorial theory of gravitation which is known as
general relativity (GR). GR theory has been accepted as a
principal method to define geometrical characteristics of
space-time. Black holes and existence of gravitational waves are
the most considerable anticipations of GR theory. Recent
observations of gravitational waves from a binary black hole
merger in LIGO and Virgo collaboration provided a firmly evidence
of Einstein's theory \cite{PRL}. In order to have an effective GR
theory, one needs the language of geometry. Hence, the approach
one prefers is to concentrate on the technique of geometry without
taking into account the complicated algebraic significances.

On the other hand, the coupling of nonlinear sources and GR, attracted
significant attentions because of their specific properties. Interesting
properties of various nonlinear electrodynamics have been studied before by
many authors \cite%
{Nonlie1,Nonlie2,Nonlie3,Nonlie4,Nonlie5,Nonlie6,Nonlie7,Nonlie8,
Nonlie9,Nonlie10,Nonlie11,Nonlie13,Nonlie14,Nonlie15,Nonlie16,
Nonlie17,Nonlie18,Nonlie19,Nonlie20,Nonlie21,Nonlie22,Nonlie23,
Nonlie24,Nonlie25}. One of the special classes of the nonlinear
electrodynamic sources is power-law Maxwell invariant (PMI), which
its Lagrangian is an arbitrary power of Maxwell Lagrangian
\cite{PM1,PM2,PM3,PM4,PM7}. It is notable that, this Lagrangian is
invariant under the conformal transformation $g_{\mu \nu
}\longrightarrow \Omega ^{2}g_{\mu \nu }$ and $A_{\mu
}\longrightarrow A_{\mu }$, where $g_{\mu \nu } $ and $A_{\mu }$
are metric tensor and the electromagnetic potential, respectively.
This model is considerably richer than Maxwell theory and in a
special case (unit power), it reduces to linear Maxwell field (see
refs. \cite{PM1,PM2,PM3,PM4,PM7}, for more details). The studies
on the black object solutions coupled to the PMI field have got a
lot of attentions in the past decade \cite
{Blackobj1,Blackobj2,Blackobj22,Blackobj3,Blackobj4,Blackobj5}.

Another attractive feature of the PMI theory is its conformal invariance,
when the power of Maxwell invariant is a quarter of space-time dimensions ($%
(n+1)/4$, where $(n+1)$ is related to dimensions of space-time). In other
words, for the special choice of $power = dimensions/4$ ($s=(n+1)/4$, where $%
s$ is power of PMI), one obtains traceless energy-momentum tensor
which leads to conformal invariancy. It is notable that the idea
is to take advantages of the conformal symmetry to construct the
analogues of $4$-dimensional Reissner-Nordstr\"{o}m solutions with
an inverse square electric field in arbitrary dimensions (see
refs. \cite{PM7,PMI1,PMI2,PMI3,PMI4}, for more details).

In recent years, more attentions are oriented on the black hole
physics, specially its critical behavior and phase transition.
Besides, interesting consequences of AdS/CFT correspondence can
motivate one for investigating the anti de-sitter space-time.
Thermodynamical behavior of the black holes in asymptotically anti
de-Sitter space-time was studied first by Hawking and Page
\cite{HP1,HP2,HP3,HP4}. Generally, at the critical point where
phase transition occurs, discontinuity/divergency of a state space
variable such as heat capacity is observed \cite{HC}.

Recently, the consideration of cosmological constant as a
thermodynamical variable, pressure, opened up new avenues in
studying black holes. It was possible to obtain van der Waals like
behavior and the second order phase transition in thermodynamical
phase structure of the black holes. In addition, it was shown that
for specific black holes, reentrant of
the phase transition and existence of triple point could be observed \cite%
{CosmP1,CosmP2,CosmP3,CosmP4a,CosmP4b,CosmP4c,CosmP5,CosmP6,CosmP7,
CosmP8,CosmP9,CosmP10,CosmP11,CosmP12,CosmP13,CosmP15,
CosmP16,CosmP17,CosmP18,CosmP19,CosmP20,CosmP21,CosmP22,CosmP23}.
The consideration of cosmological constant as a thermodynamical
variable has been supported by studies that are conducted in the
context of the classical thermodynamics of black holes and
gauge/gravity duality
\cite{CosVar2,CosVar3,CosVar4,CosVar5,Mamasani}. One of the
methods for obtaining van der Waals like critical points is
through the use of heat capacity with a relation between the
cosmological constant and pressure \cite{int}. This method has
been employed in several papers and it was shown that its results
are consistent with those obtained through regular methods.

On the other hand, geometrical thermodynamics (GT), which was
first employed by Gibbs and Caratheodory, is another interesting
way for investigation of black hole phase transition. Regarding
this method, one could build a phase space by employing
thermodynamical potential and its corresponding extensive
parameters. Meanwhile, divergence points of thermodynamical Ricci
scalar provide information related to thermodynamic phase
transition points. Historically speaking, first, Weinhold
introduced a metric on the equilibrium thermodynamical phase space
\cite{WeinholdI,WeinholdII} and after that, it was redefined by
Ruppeiner from a different point of view
\cite{RuppeinerI,RuppeinerII}. It is worthwhile to mention that,
there is a conformally relationship between Ruppeiner/Weinhold
metric. One can obtain that, the conformal factor is inverse of
temperature \cite{Salamon}. None of Weinhold and Ruppeiner metrics
were invariant under Legendre transformation.
The first Legendre invariance metric was introduced by Quevedo \cite%
{QuevedoI,QuevedoII} which can remove some problems of
Weinhold/Ruppeiner methods. Although Quevedo metric can solve some
issues of the previous methods, it is confronted with other
problems in specific systems. Therefore, in order
to remove these problems, a new method was proposed in Ref. \cite%
{HPEMI,HPEMII,HPEMIII,HPEMIV} which is known as HPEM
(Hendi-Panahiyan-Eslam Panah-Momennia) metric. In this paper, we
want to study thermal stability and phase transition in the
context of GT method and extended phase space for black holes in
Einstein gravity with the PMI source in higher dimensions. It is
notable that in recent years, phase transition, curvatures,
Hessian
matrix and Nambu brackets have been studied via a new GT method \cite%
{Mirza1,Mirza2,Mirza3,Roy} which, here, we are not interested to
discuss it.

The structure of this paper is as follows. First, we introduce the
field equations and black hole solutions in Einstein-PMI gravity.
Next, temperature and heat capacity for the obtained black hole
solution will be investigated and the stability will be studied.
Then, Weinhold, Ruppeiner and Quevedo metrics for studying
geometrical thermodynamics of these black holes are employed. It
will be seen that, these metrics fail to provide fruitful results.
So, we employ the HPEM metric and study the phase transition of
these black holes in the context of GT. Next, the critical points
are extracted through the use of proportionality between the
cosmological constant and pressure. Finally, we finish our paper
with some closing remarks.


\section{FIELD EQUATIONS AND SOLUTIONS}

The action of $(n+1)$-dimensional Einstein gravity in the presence
of power-law Maxwell field with the negative cosmological constant
can be written as
\begin{equation}
I=-\frac{1}{16\pi }\int\limits_{M}d^{n+1}x\sqrt{-g}\left[ R-2\Lambda
+(-F)^{s}\right] ,  \label{Eaction}
\end{equation}%
where $R$ is the Ricci scalar and $s$ is constant which
determining the nonlinearity power of electromagnetic field. The
Maxwell invariant $F=F_{\mu \nu }F^{\mu \nu }$, where $F_{\mu \nu
}$ is the electromagnetic tensor which is equal to $\partial _{\mu
}A_{\nu }-\partial _{\nu }A_{\mu }$ and $A_{\mu } $ is the gauge
potential one-form. One can obtain the field equations by varying
action (\ref{Eaction}) with respect to the gravitational and gauge
field, $g_{\mu \nu }$ and $A_{\mu }$, respectively
\begin{eqnarray}
G_{\mu \nu }+\Lambda g_{\mu \nu } &=&T_{\mu \nu },  \label{Efield} \\
\partial _{\mu }\left[ \sqrt{-g}(-F)^{s-1}F^{\mu \nu }\right] &=&0.
\label{Eelectro}
\end{eqnarray}

In the presence of nonlinear power-law Maxwell invariant field, one can show
that the energy-momentum tensor becomes
\begin{equation}
T_{\mu \nu }=\left[ \frac{1}{2}g_{\mu \nu }(-F)^{s}+2sF_{\mu \sigma }F_{\nu
}^{\sigma }(-F)^{s-1}\right] ,  \label{energy}
\end{equation}%
and for the case of $s=1$ Eqs. (\ref{Efield})-(\ref{energy}) reduce to
well-known Einstein-Maxwell theory \cite{Max-Eins}. We apply the following
static metric of $(n+1)$-dimensional space-time
\begin{equation}
ds^{2}=-W(r)dt^{2}+ \frac{dr^{2}}{W(r)}+{r^{2}}d\Omega _{n-1}^{2} ,
\label{metric}
\end{equation}%
where $W(r)$ is an arbitrary function of radial coordinate which should be
determined, and $d\Omega _{n-1}^{2}$ is the line element of $(n-1)$%
-dimensional hypersurface with volume $\omega _{n-1}$ which has constant
curvature $(n-1)(n-2)k$
\begin{equation}
d\Omega _{n-1}^{2}=\left\{
\begin{array}{cc}
d\theta _{1}^{2}+\sum\limits_{i=2}^{n-1}\prod\limits_{j=1}^{i-1}\sin
^{2}\theta _{j}d\theta _{i}^{2} & k=1 \\
d\theta _{1}^{2}+\sinh ^{2}\theta _{1}\left[ d\theta
_{2}^{2}+\sum\limits_{i=3}^{n-1}\prod\limits_{j=2}^{i-1}\sin ^{2}\theta
_{j}d\theta _{i}^{2}\right] & k=-1 \\
\sum\limits_{i=1}^{n-1}d\phi _{i}^{2} & k=0%
\end{array}%
\right. . \label{metrick}
\end{equation}

Regarding the constant $k$, one finds that the boundary of
$t=constant$ and $r=constant$ can be a positive (spherical),
negative (hyperbolic) and zero (flat) constant curvature
hypersurface. Since we are going to obtain the electrically
charged solutions, we consider the consistent gauge potential
one-form as $A=h(r)dt$. By considering this radial gauge
potential, we find that the Maxwell invariant will be $F=-\left(
\frac{dh(r)}{dr}\right) ^{2}$, and therefore, regardless of the
values of nonlinearity parameter ($s$), the solutions are
well-defined. Now, taking into account the mentioned gauge
potential with Eqs. (\ref{Efield}) and (\ref{Eelectro}), one can
obtain the metric function as well as the electromagnetic field in
following forms \cite{CosmP9}
\begin{equation}
W(r)=k-\frac{2\Lambda r^{2}}{n(n-1)}-\frac{m}{r^{n-2}}+\left\{
\begin{array}{cc}
\begin{array}{c}
-\frac{2^{n/2}q^{n}}{r^{n-2}}\ln \left( \frac{r}{l}\right) \\
\\
\end{array}
&
\begin{array}{c}
s=\frac{n}{2} \\
\\
\end{array}
\\
\frac{(2s-1)^{2}\left( \frac{(n-1)(2s-n)^{2}q^{2}}{(n-2)(2s-1)^{2}}\right)
^{s}}{r^{2(ns-3s+1)/(2s-1)}(n-1)(n-2s)} & otherwise%
\end{array}%
\right. ,  \label{W(r)}
\end{equation}%
\begin{equation}
F_{tr}=\frac{dh(r)}{dr}=\left\{
\begin{array}{cc}
\begin{array}{c}
-\frac{q}{r} \\
\\
\end{array}
&
\begin{array}{c}
s=\frac{n}{2} \\
\\
\end{array}
\\
-q\left( \frac{2s-n}{2s-1}\right) \sqrt{\frac{\left( n-1\right) }{2(n-2)}}%
r^{-\left( \frac{n-1}{2s-1}\right) } & otherwise%
\end{array}%
\right. ,  \label{E(r)}
\end{equation}%
where $q$ and $m$ are integration constants which are related to electric
charge ($Q$) and the ADM mass ($M$) of the black hole in the following
manner
\begin{equation}
M=\frac{(n-1)}{16\pi }m,  \label{M}
\end{equation}%
\begin{equation}
Q=\left\{
\begin{array}{cc}
\begin{array}{c}
\frac{n}{2\pi }2^{\frac{n-6}{2}}q^{n-1} \\
\\
\end{array}
&
\begin{array}{c}
s=\frac{n}{2} \\
\\
\end{array}
\\
\frac{\sqrt{2}(2s-1)s}{8\pi }\left( \frac{n-1}{n-2}\right) ^{s-1/2}\left(
\frac{(n-2s)q}{2s-1}\right) ^{2s-1} & otherwise%
\end{array}%
\right. .  \label{Q}
\end{equation}

By using the area law, the black hole entropy could be determined
as
\begin{equation}
S=\frac{r_{+}^{n-1}}{4},  \label{S}
\end{equation}
where $r_{+}$ is the radius of horizon (event horizon). We can
also obtain the Hawking temperature of black hole on the outer
(event) horizon by using surface gravity interpretation
\begin{eqnarray}
T &=&\frac{1}{2\pi }\sqrt{-\frac{1}{2}\left( \nabla _{\mu }\chi _{\nu
}\right) \left( \nabla ^{\mu }\chi ^{\nu }\right) }=\frac{W^{\prime }(r_{+})%
}{4\pi }  \notag \\
&=&\frac{\left( n-2\right) k}{4\pi r_{+}}-\frac{r_{+}\Lambda }{2\pi (n-1)}%
-\left\{
\begin{array}{cc}
\begin{array}{c}
\frac{2^{\frac{n-4}{2}}q^{n}}{\pi r_{+}^{n-1}} \\
\end{array}
&
\begin{array}{c}
s=\frac{n}{2} \\
\end{array}
\\
\frac{\left( 2s-1\right) }{4\pi (n-1)}\frac{\left( \frac{(n-1)(2s-n)^{2}q^{2}%
}{(n-2)(2s-1)^{2}}\right) ^{s}}{r_{+}^{(2s\left( n-2\right) +1)/(2s-1)}} &
otherwise%
\end{array}%
\right. ,  \label{T}
\end{eqnarray}%
where $\chi=\partial_{t} $ is the Killing vector of the event
horizon. Also, the black hole electric potential ($U$), could be
measured at infinity with respect to the event horizon $r_{+}$
\begin{equation}
U=\left\{
\begin{array}{cc}
\begin{array}{c}
-q\ln \left( \frac{r_{+}}{l}\right) \\
\\
\end{array}
&
\begin{array}{c}
s=\frac{n}{2} \\
\\
\end{array}
\\
\sqrt{\frac{\left( n-1\right) }{2(n-2)}}\frac{q}{r_{+}^{(n-2s)/(2s-1)}} &
otherwise%
\end{array}%
\right. .  \label{U}
\end{equation}

It is straightforward to show that these quantities satisfy the first law of
thermodynamics,
\begin{equation}
dM=TdS+UdQ.  \label{First-Law}
\end{equation}

The case where $s=\frac{n}{2}$ is known as conformally invariant
Maxwell (CIM). Through the paper, we will divide the solutions to
the cases of CIM (where $s=\frac{n}{2}$) and PMI ($s\neq
\frac{n}{2}$) solutions. In the following, by using the
geometrical thermodynamics approach, we want to study the phase
transitions of black hole.

\section{ thermodynamical structure}

\subsection{Temperature}

\subsubsection{PMI case}

In the case of PMI, it is evident that signature of the
temperature depends on choices of different parameters. Here, for
the AdS spherical case, the topological ($\kappa$) and
cosmological constant ($\Lambda$) terms are contributing to
positivity of the temperature. Interestingly for the charge term,
depending on choices of nonlinearity parameter, $s$, this term may
contribute to the negativity or positivity of temperature. For
$s>1/2$, the charge term will always have negative effect on the
temperature while for violation of this condition, it will have a
positive contribution. It is crucial to mention that this dual
behavior is due to contribution of the generalization to nonlinear
electromagnetic field. In other words, by setting $s=1$, the
charge term will only have negative effect on the temperature
(contribution to positivity does not exist). This highlights the
effects of nonlinear electromagnetic field on the thermodynamical
structure of black holes. That being said, one can point out that
for violation of the mentioned condition, for spherically
symmetric AdS black holes, temperature will be positive without
any root.

For small values of the horizon radius, dominant term is the
charge term while for the large values of horizon radius, $\Lambda
$ term will be dominant. For spherically symmetric AdS space-time,
$\Lambda $ term is positive. Therefore, for the large values of
horizon radius, temperature will be positive. Therefore, there
exists a root for the temperature, $r_{0}$, in which for
$r_{+}<r_{0}$, the temperature is negative and solutions are not
physical.

On the contrary, due to negative contribution of the $\Lambda $ in
spherical dS space-time, the temperature will be negative for
large values of horizon radius. The effective term here is
topological term. In other words, by suitable choices of different
parameters, for a region of horizon radius, topological term would
be dominant which results into formation of an extremum (maximum)
and existence of two roots. Between these two roots, temperature
is positive and solutions are physical (otherwise, temperature is
negative). It is worthwhile to mention that for hyperbolic
horizon, the temperature will be always negative in this case.
Therefore, there is no any physical solution in this case.

Now, we focus on effects of each parameter on negativity/positivity of the
temperature. Evidently, by increasing dimensions, the effect of topological
and charge terms increase while the opposite takes place for $\Lambda $. As
for nonlinearity parameter, for $r_{+}>1$, if $0.5<s<1$, then the effects of
charge term is an increasing function of $s$, while if $1<s$, the effects of
charge term will be a decreasing function of $s$. It is worthwhile to
mention that for $r_{+}<1,$ for both cases of $0.5<s<1$ and $1<s$, the
charge term is a decreasing function of the nonlinearity parameter. We
should point it out that we have excluded $s=1$, since it is Maxwell case.
In addition, the effect of charge term is an increasing function of the
electric charge. Now, by considering mentioned effects, one is able to
determine the thermodynamical behavior of these black holes in the context
of temperature and study different limits which these black holes have.

\subsubsection{CIM case}

In the CIM case, contribution of the charged term is always toward
negativity of the temperature. The dominant term for the small
values of horizon radius is the charge term which is negative. For
the medium and large values of horizon radius, the dominant terms,
respectively, are topological and $\Lambda $ terms. Considering
the dominance of different terms, depending on choices of topology
and type of space-time, the temperature could be one of the
following cases:

I) For dS space-time and $k=0$, $-1$, all the terms in temperature are
negative. Therefore, temperature will be negative and solutions are not
physical.

II) For dS space-time and $k=1$, only the topological term is
positive whereas the charge and $\Lambda $ are negative.
Remembering that for small and large black holes, the dominant
term are $q$ and $\Lambda $ terms, it is possible to find two
roots for temperature and one maximum which is located at the
positive values of temperature. This leads to presence of physical
solutions only for medium black holes while for large and small
black holes, physical black holes are absent.

III) For AdS space-time, the $\Lambda $ term has positive
contribution. Remembering that for small and large black holes,
dominant terms are $q$ and $\Lambda $ terms, respectively,
irrespective of topological structure of the black holes, there
exists a root for the temperature. For black holes smaller than
this root, temperature is negative and solutions are non-physical.
For $k=0$, $-1$, the temperature is only an increasing function of
the horizon radius. Whereas, by suitable choices of different
parameters, in case of $k=1$, temperature may acquire one or two
extrema. The behavior of temperature in this case is similar to
$T-r_{+}$ diagrams in extended phase space (similar ones in van
der Waals like black holes), which indicates that a second order
phase transition takes place for these black holes. It is
worthwhile to mention that extrema in temperature are matched with
divergencies in the heat capacity which results into the existence
of second order phase transition in thermodynamical structure of
the black holes.

\subsection{Heat capacity and stability}

Next, we study the stability conditions of these black holes. To
do so, we employ the canonical ensemble which is based on the heat
capacity. The stability conditions are determined by the behavior
of heat capacity. In other words, the sign of heat capacity
represents thermal stability/instability of the system. The
positivity indicates that system under consideration is in
thermally stable state. In addition, there are two types of points
which could be extracted by using the heat capacity: bound and
phase transition points. The bound point is where the heat
capacity (temperature) acquires a root. The reason for calling it
bound point comes from the fact that it is where the sign of
temperature is changed. Since the negative temperature is
representing a non-physical solution, this point marks a bound
point. On the other hand, phase transition point is where a
discontinuity exists for the heat capacity.

The heat capacity is given by
\begin{equation}
C_{Q}=\frac{\left( \frac{\partial M}{\partial S}\right) _{Q}}{\left( \frac{%
\partial ^{2}M}{\partial S^{2}}\right) _{Q}}=T\left( \frac{\partial S}{%
\partial T}\right) _{Q},  \label{heat capacity}
\end{equation}%
where by employing Eqs. (\ref{S}) and (\ref{T}), one can find
\begin{equation}
C_{Q}=\left\{
\begin{array}{cc}
\begin{array}{c}
-\frac{(n-1)(n-2)kr_{+}^{2n-3}-2\Lambda r_{+}^{2n-1}-2^{\frac{n}{2}}\left(
n-1\right) q^{n}r_{+}^{n-1}}{4\left[ (n-2)kr_{+}^{n-2}+\frac{2\Lambda }{(n-1)%
}r_{+}^{n}-2^{\frac{n}{2}}\left( n-1\right) q^{n}\right] } \\
\\
\end{array}
&
\begin{array}{c}
s=\frac{n}{2} \\
\\
\end{array}
\\
-\frac{\left( n-1\right) (n-2)kr_{+}^{\frac{4ns-n+2}{2s-1}}-2\Lambda r_{+}^{%
\frac{4s\left( n+1\right) -n}{2s-1}}-(2s-1)\left( \frac{(n-1)(n-2s)^{2}q^{2}%
}{(n-2)(2s-1)^{2}}\right) ^{s}r_{+}^{\frac{6s+n\left( 2s-1\right) }{2s-1}}}{4%
\left[ (n-2)kr_{+}^{\frac{2(ns+1)}{2s-1}}+\frac{2\Lambda }{(n-1)}r_{+}^{%
\frac{2s(n+2)}{2s-1}}-\frac{(2s\left( n-2\right) +1)}{(n-1)}\left( \frac{%
(n-1)(n-2s)^{2}q^{2}}{(n-2)(2s-1)^{2}}\right) ^{s}r_{+}^{\frac{6s}{2s-1}}%
\right] r_{+}} & otherwise%
\end{array}%
\right. .  \label{Heat}
\end{equation}

\subsubsection{PMI case}

The positivity of heat capacity is determined by the signature of
denominator and numerator of it. Here, in order to have a positive
heat capacity, two different cases could be considered; whether
numerator ($A$) and denominator ($B$) are positive or both of them
are negative ($A\times
B>0 $);%
\begin{eqnarray*}
A &=&2\Lambda r_{+}^{\frac{4s\left( n+1\right) -n}{2s-1}}+(2s-1)\left( \frac{%
(n-1)(n-2s)^{2}q^{2}}{(n-2)(2s-1)^{2}}\right) ^{s}r_{+}^{\frac{6s+n\left(
2s-1\right) }{2s-1}}-(n-1)(n-2)kr_{+}^{\frac{n\left( 4s-1\right) +2}{2s-1}},
\\
&& \\
B &=&(n-2)kr_{+}^{\frac{2(ns+1)}{2s-1}}+\frac{2\Lambda }{(n-1)}r_{+}^{\frac{%
2s(n+2)}{2s-1}}-\frac{(2s\left( n-2\right) +1)}{(n-1)}\left( \frac{%
(n-1)(n-2s)^{2}q^{2}}{(n-2)(2s-1)^{2}}\right) ^{s}r_{+}^{\frac{6s}{2s-1}}.
\end{eqnarray*}

\begin{figure}[tbp]
$%
\begin{array}{ccc}
\epsfxsize=6cm \epsffile{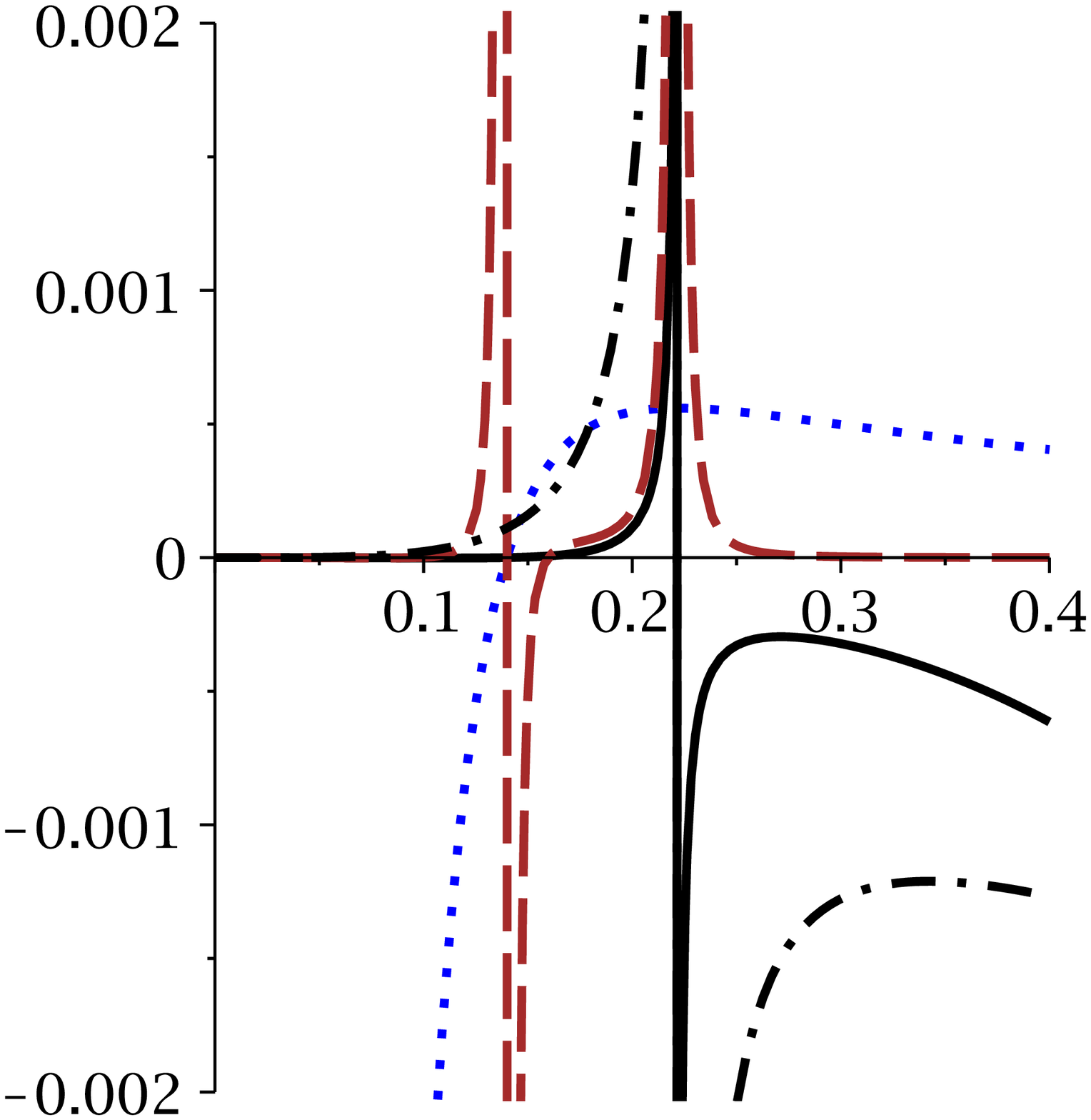} & \epsfxsize=6cm \epsffile{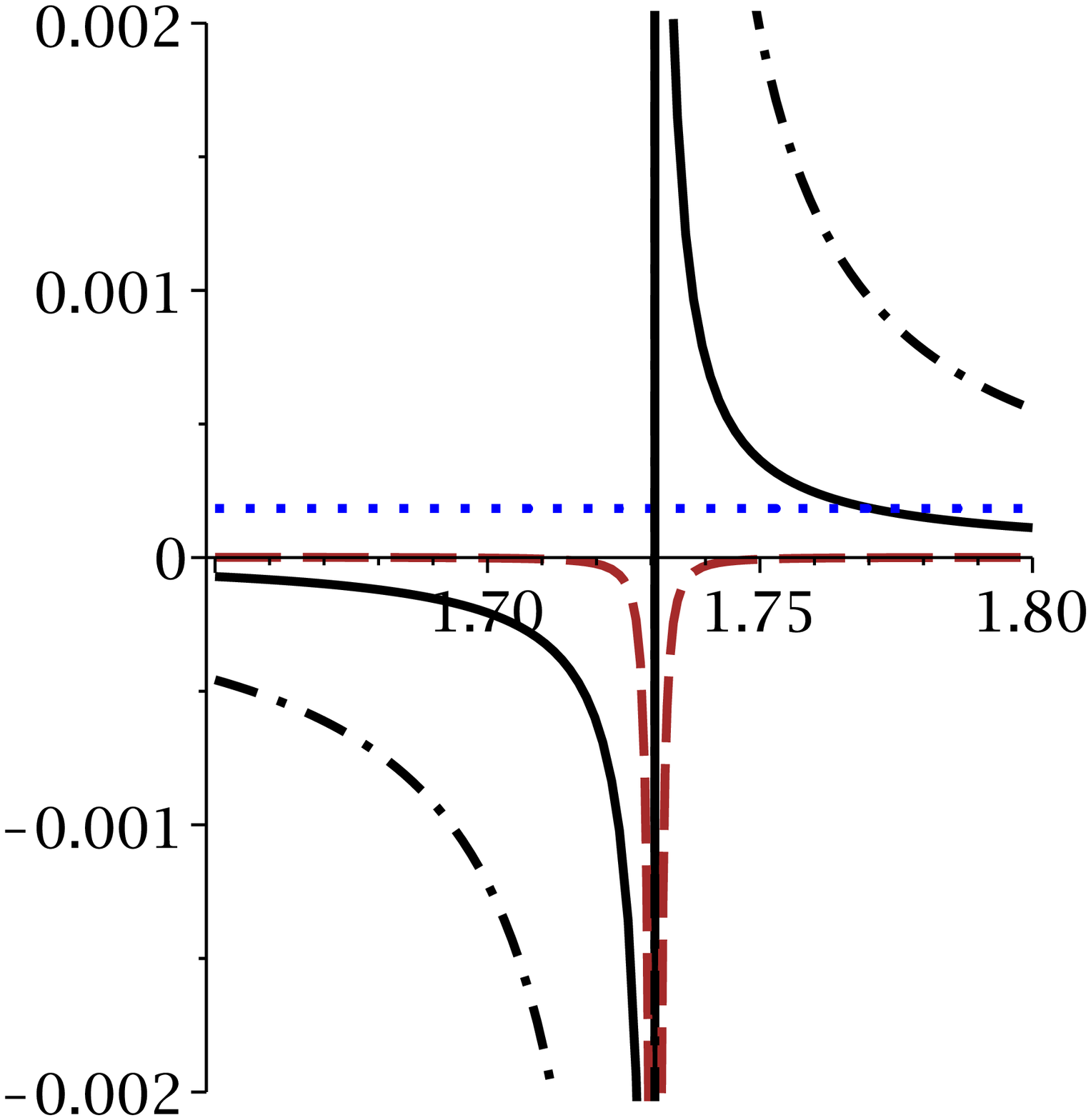} & %
\epsfxsize=6cm \epsffile{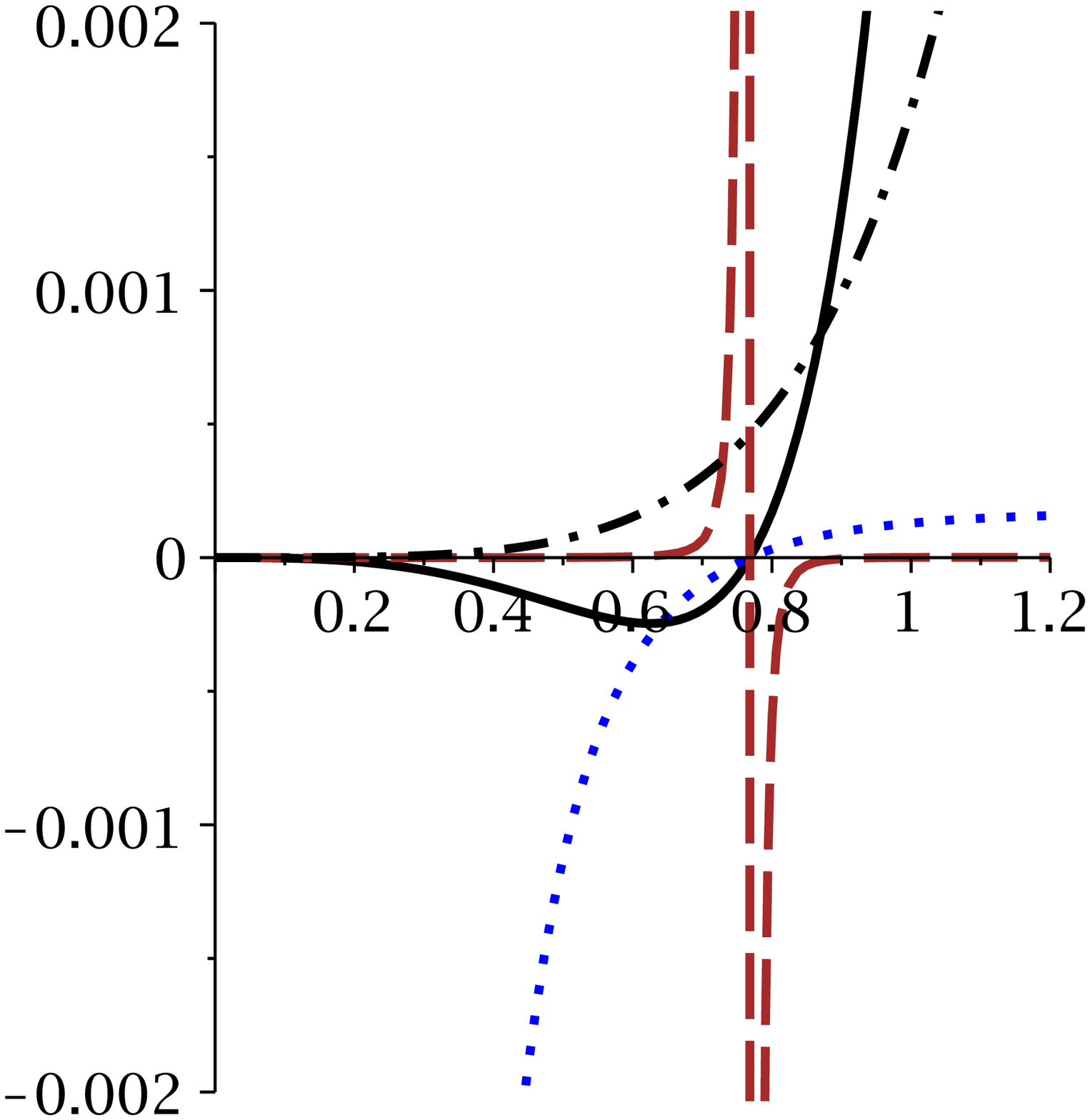}%
\end{array}
$%
\caption{\textbf{HPEM metric:} $\mathcal{R}$ (dashed line), T (dotted line),
$\protect\alpha$ (dashed-dotted line) and $C_{Q}$ (continues line) versus $%
r_{+}$ for $\Lambda=-1$, $s=1.2$ and $n=4$; Left and middle panels: $q=0.1$,
right panel: $q=1$ (\emph{For different scales}).}
\label{Fig1}
\end{figure}

\begin{figure}[tbp]
$%
\begin{array}{ccc}
\epsfxsize=6cm \epsffile{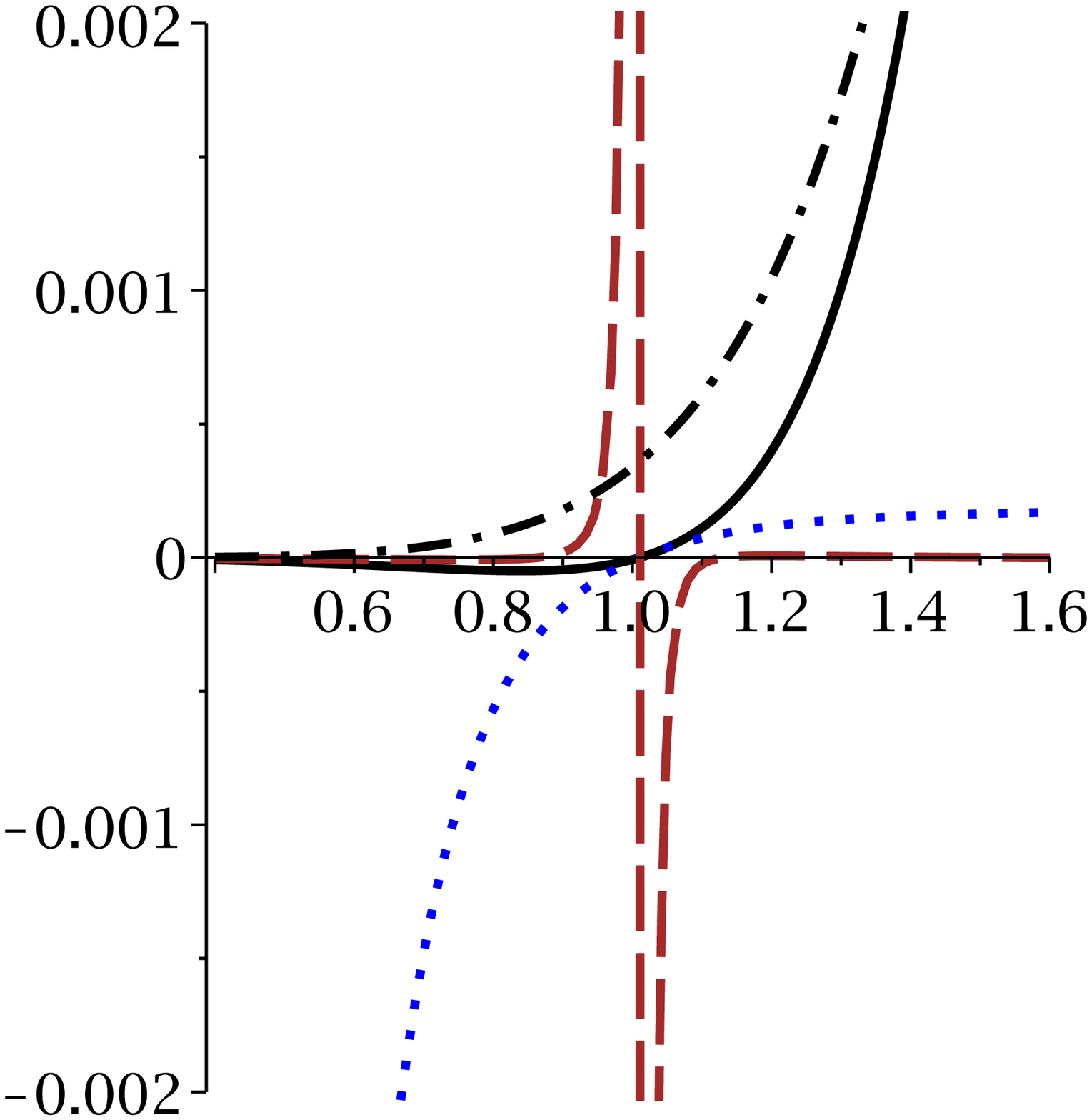} & \epsfxsize=6cm \epsffile{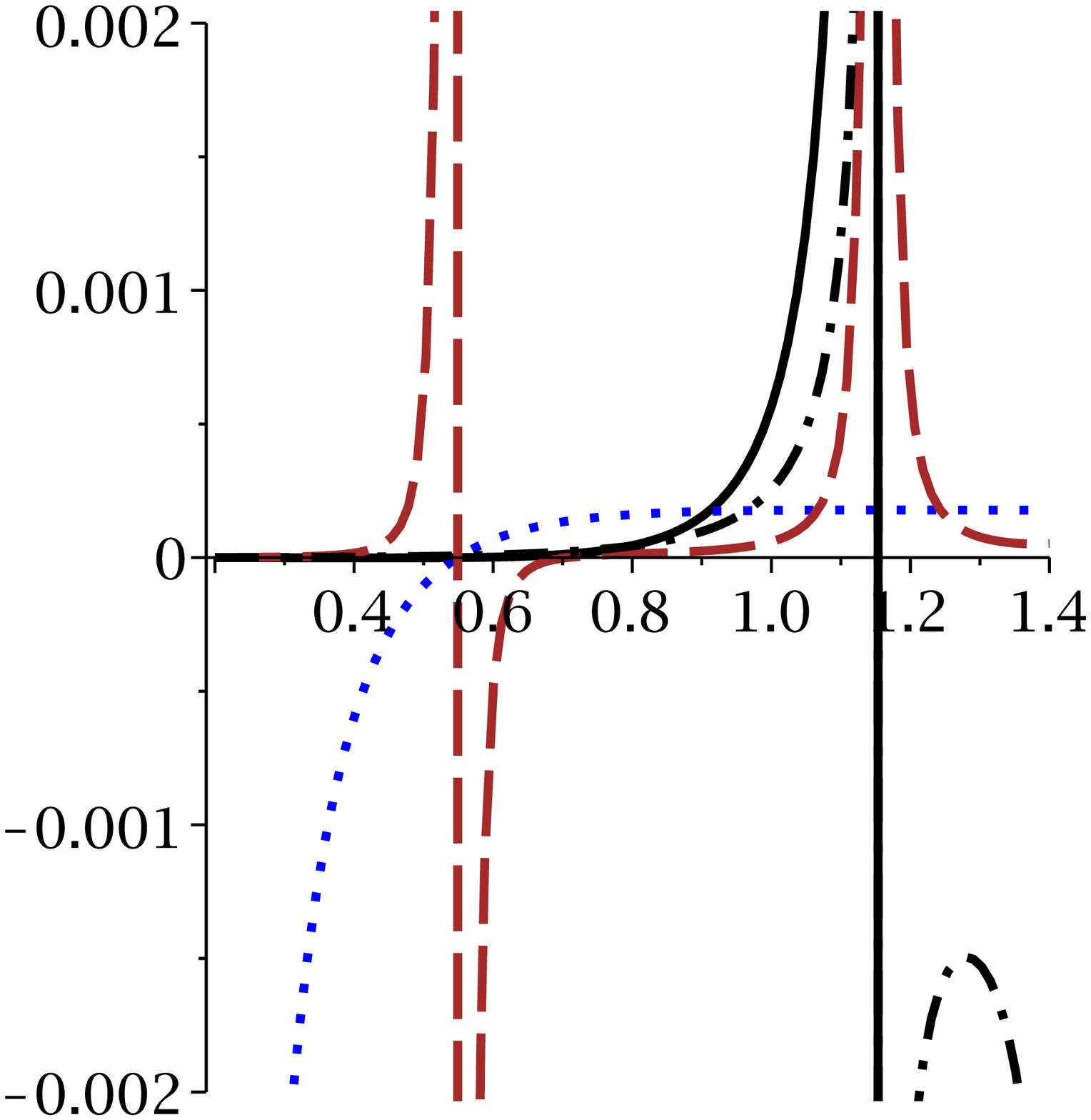} & %
\epsfxsize=6cm \epsffile{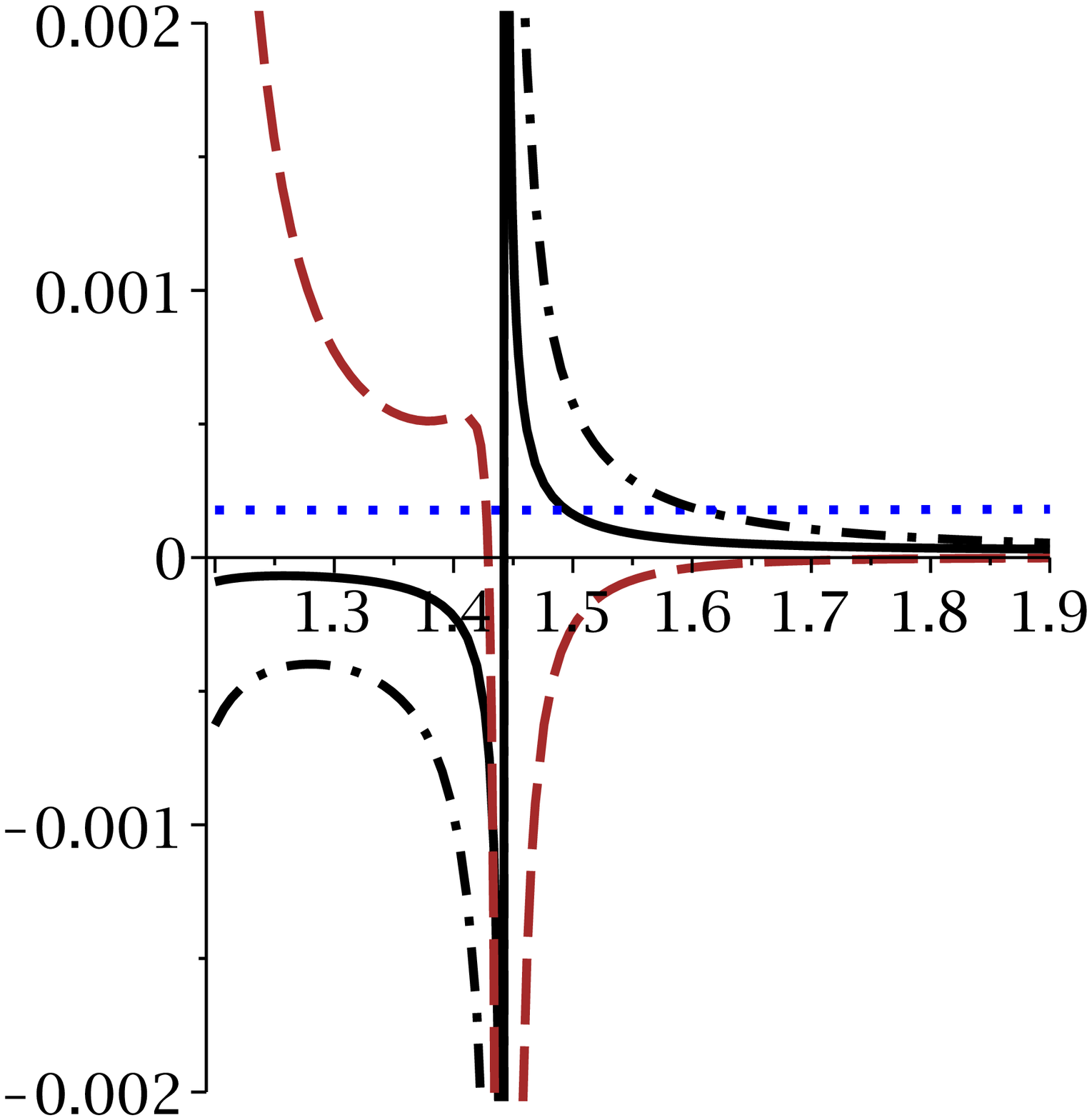}%
\end{array}
$%
\caption{\textbf{HPEM metric:} $\mathcal{R}$ (dashed line), T (dotted line),
$\protect\alpha$ (dashed-dotted line) and $C_{Q}$ (continues line) versus $%
r_{+}$ for $\Lambda=-1$, $q=1.1$ and $n=4$; Left panel: $s=0.9$, middle and
right panels: $s=1.4$ (\emph{For different scales}).}
\label{Fig2}
\end{figure}

\begin{figure}[tbp]
$%
\begin{array}{ccc}
\epsfxsize=6cm \epsffile{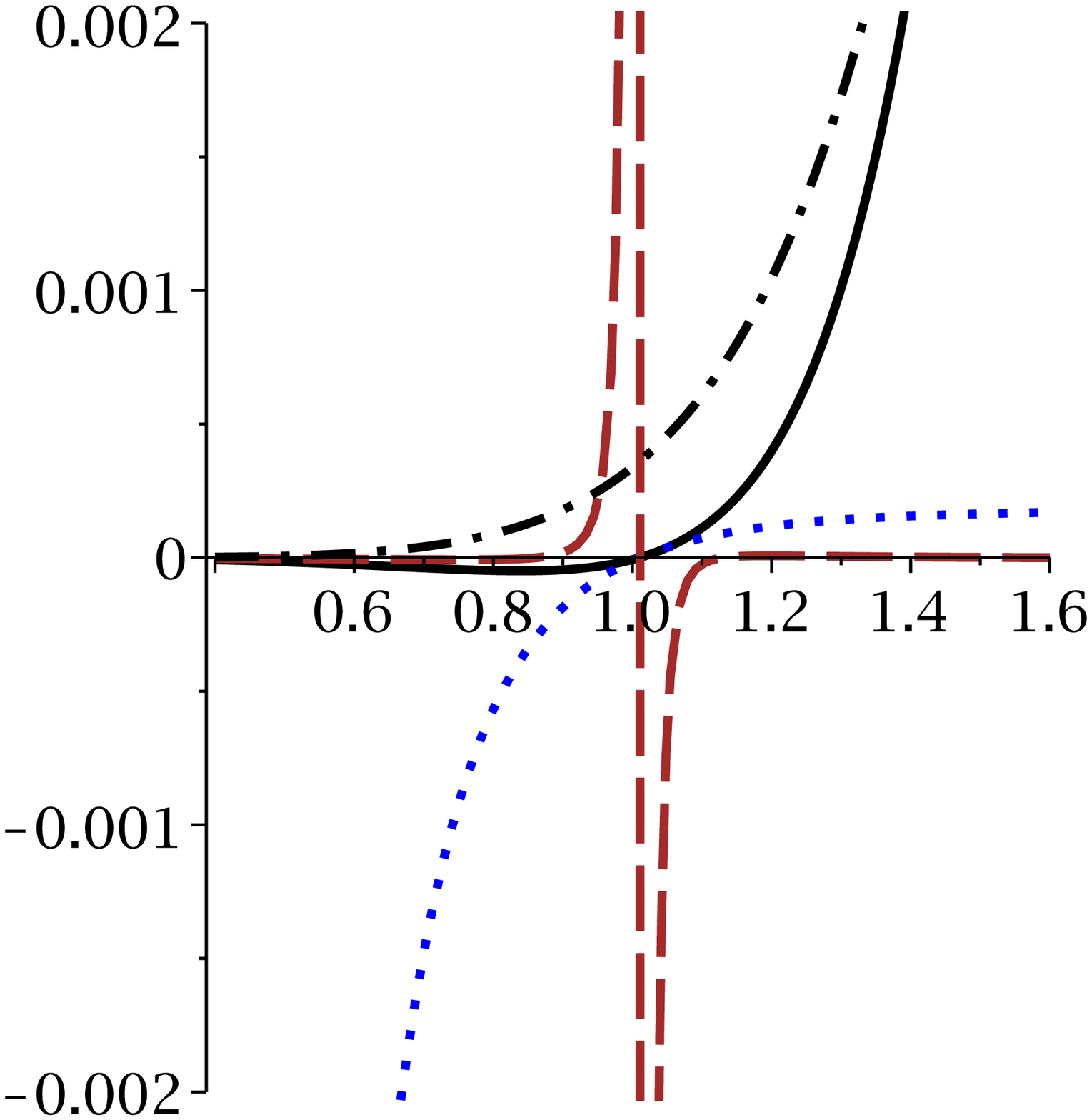} & \epsfxsize=6cm \epsffile{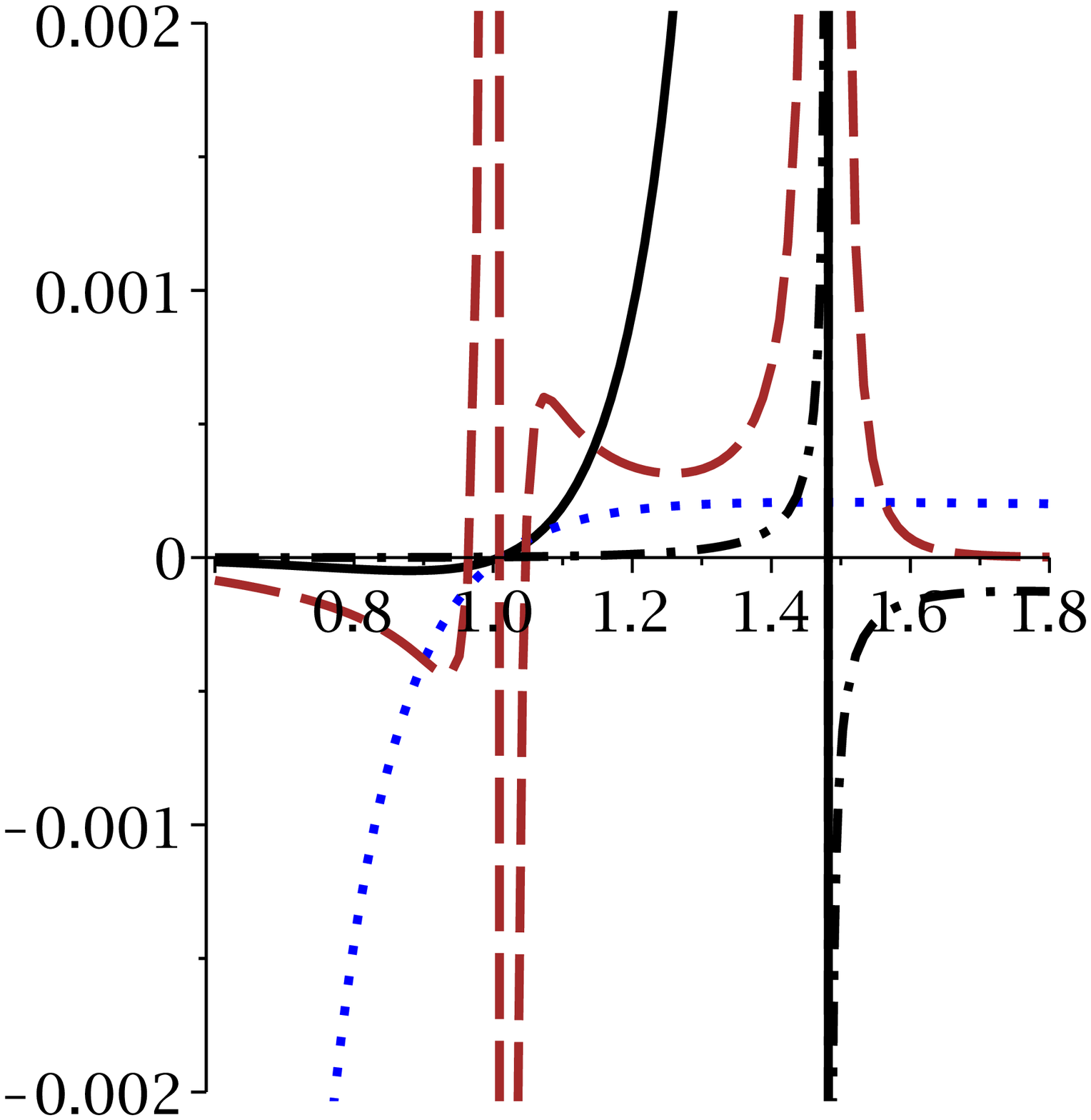} & %
\epsfxsize=6cm \epsffile{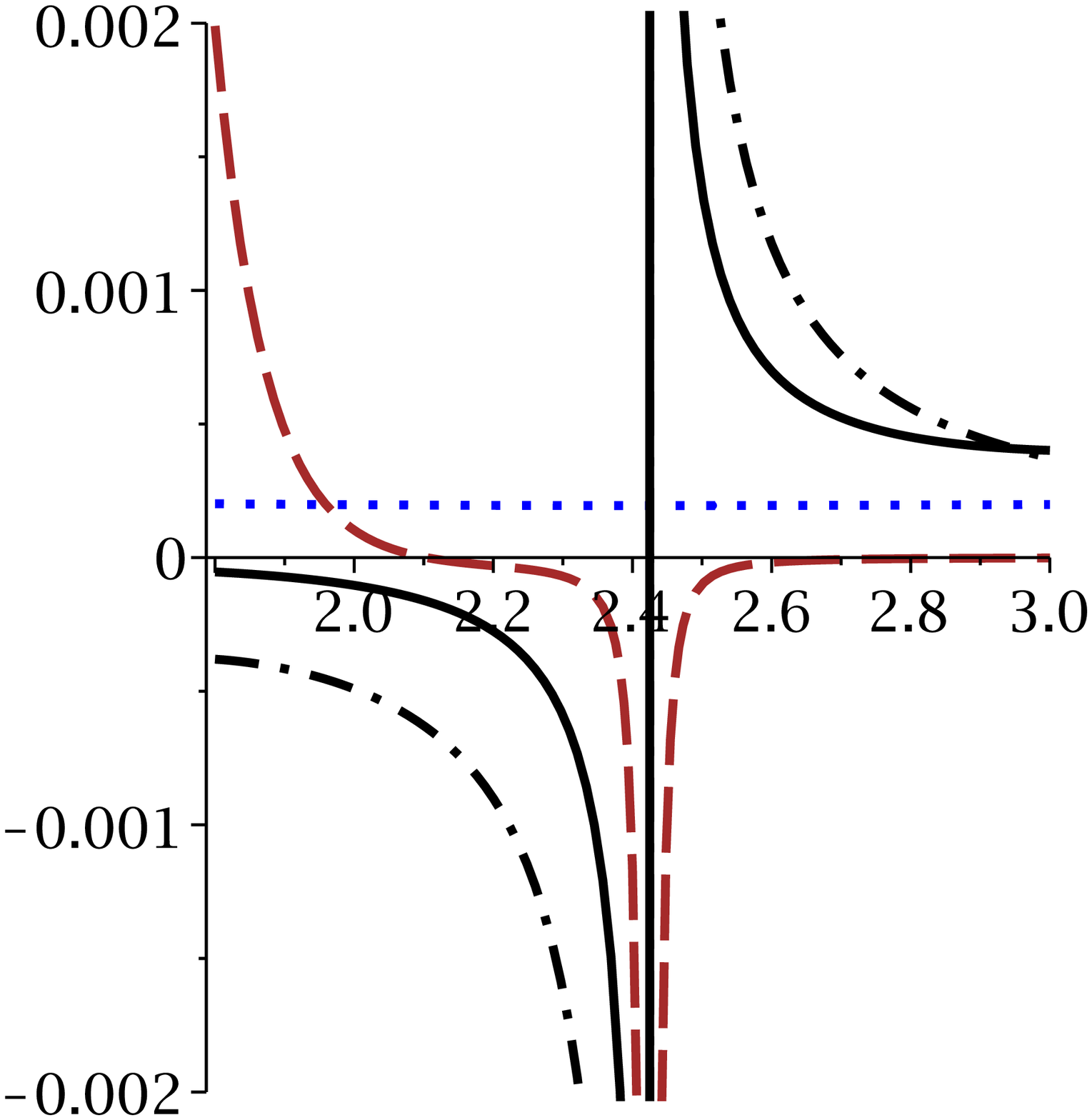}%
\end{array}
$%
\caption{\textbf{HPEM metric:} $\mathcal{R}$ (dashed line), T (dotted line),
$\protect\alpha$ (dashed-dotted line) and $C_{Q}$ (continues line) versus $%
r_{+}$ for $\Lambda=-1$, $q=1.1$ and $s=1.2$; Left panel: $n=4$, middle and
right panels: $n=5$ (\emph{For different scales}).}
\label{Fig3}
\end{figure}

Therefore, two set of conditions should be satisfied to have a
positive heat capacity, hence, thermally stable solutions. Due to
complexity of the obtained heat capacity for PMI case, it is not
possible to extract bound and phase transition points
analytically. Therefore, we employ numerical approach to study
stability and obtain bound and phase transition points of these
black holes. The results of numerical evaluation is presented in
following diagrams (Figs. \ref{Fig1}-\ref{Fig3}).

\begin{figure}[tbp]
$%
\begin{array}{ccc}
\epsfxsize=6cm \epsffile{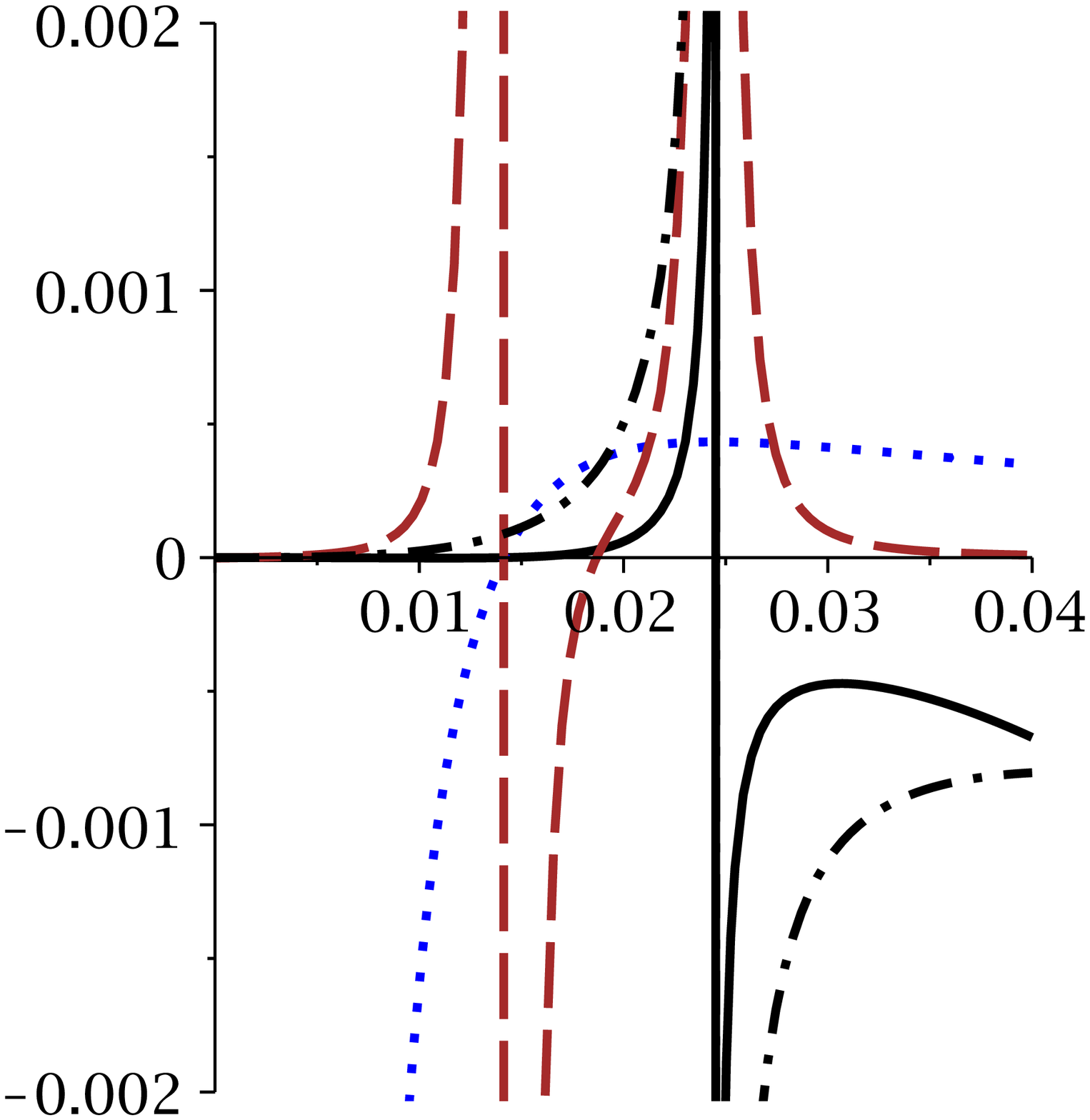} & \epsfxsize=6cm \epsffile{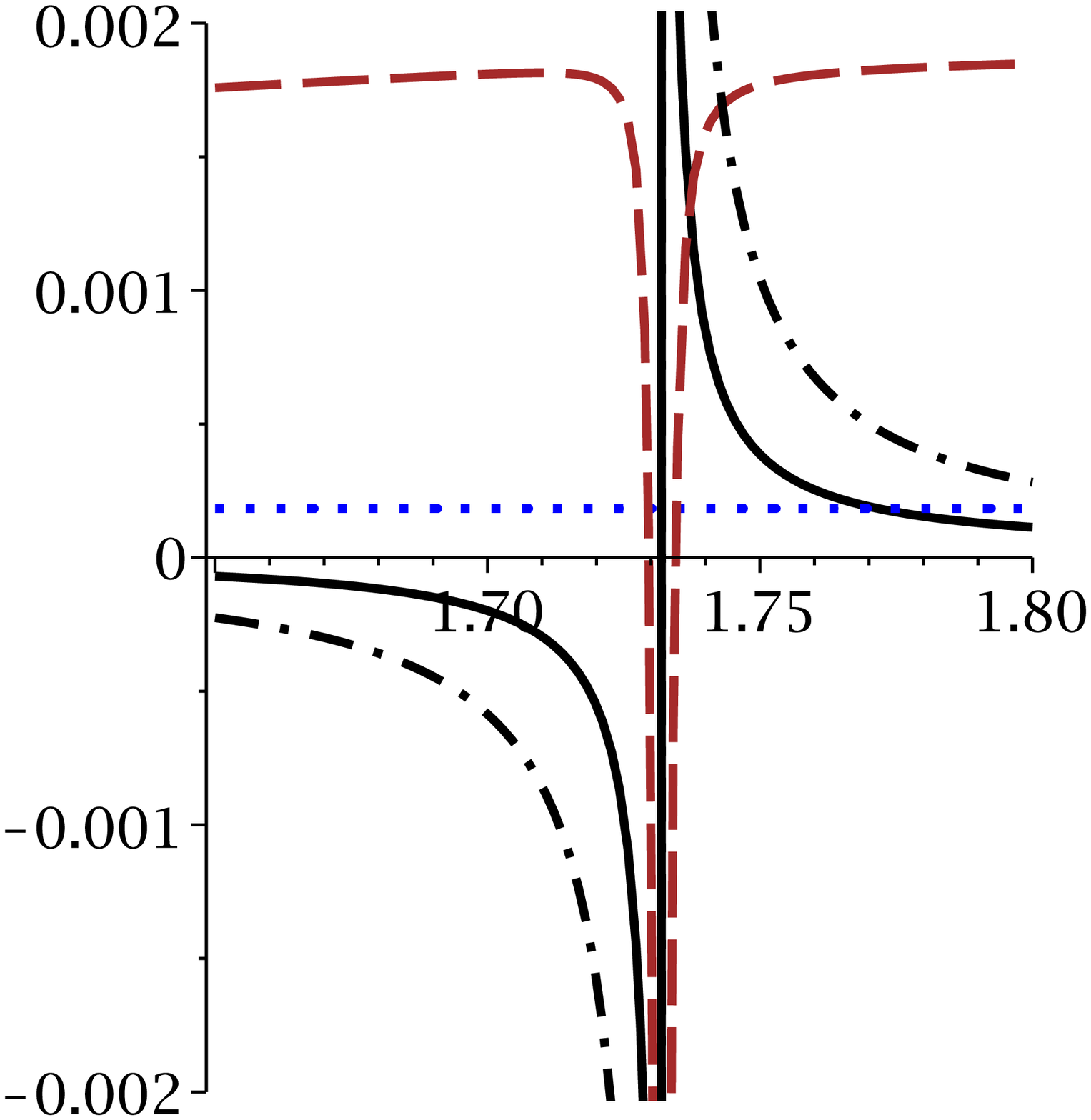} & %
\epsfxsize=6cm \epsffile{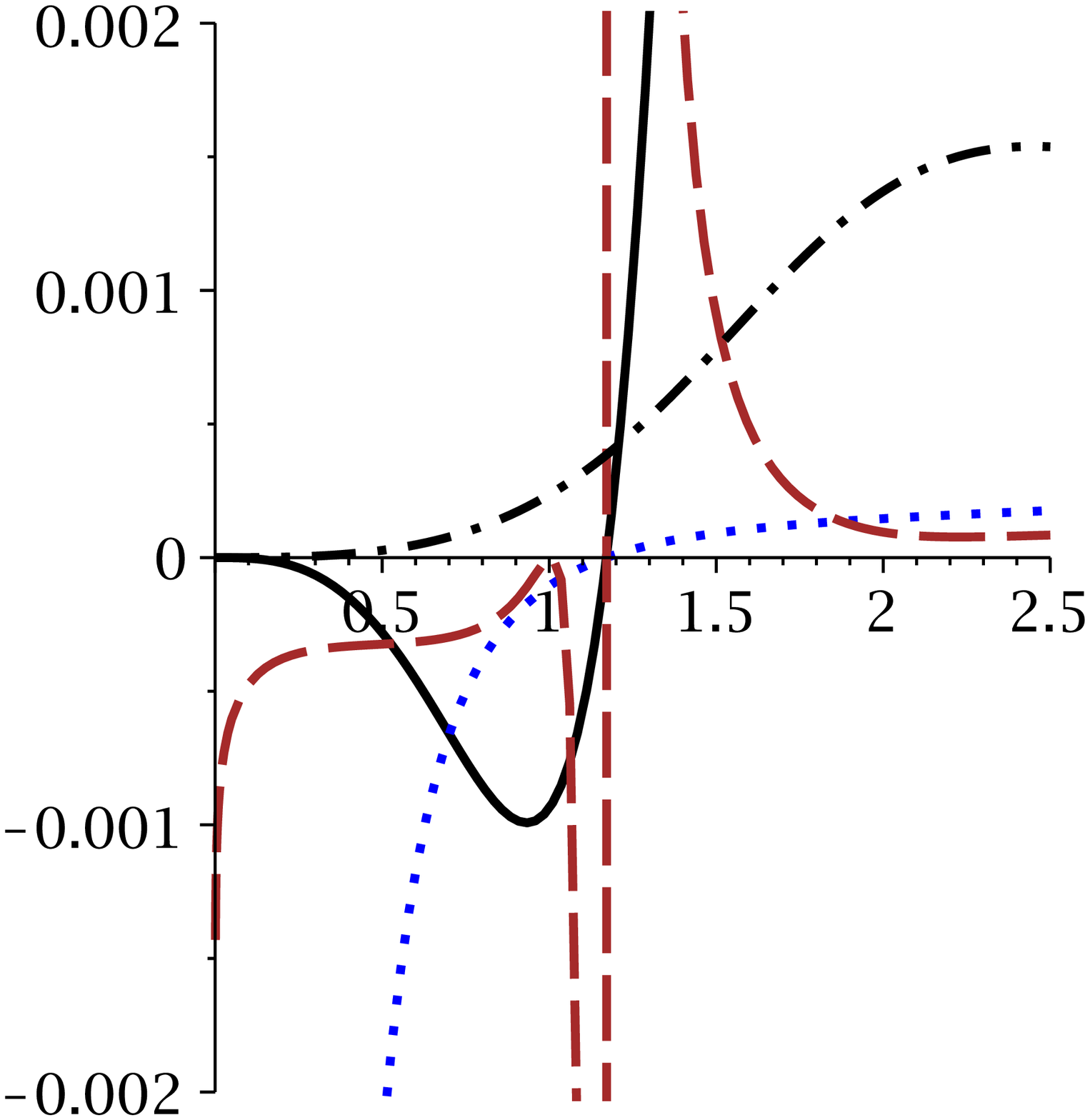}%
\end{array}
$%
\caption{\textbf{HPEM metric:} $\mathcal{R}$ (dashed line), T (dotted line),
$\protect\alpha$ (dashed-dotted line) and $C_{Q}$ (continues line) versus $%
r_{+}$ for $\Lambda=-1$, $l=1$ and $n=4$; Left and middle panels: $q=0.1$,
right panel: $q=1$ (\emph{For different scales}).}
\label{Fig4}
\end{figure}
\begin{figure}[tbp]
$%
\begin{array}{ccc}
\epsfxsize=6cm \epsffile{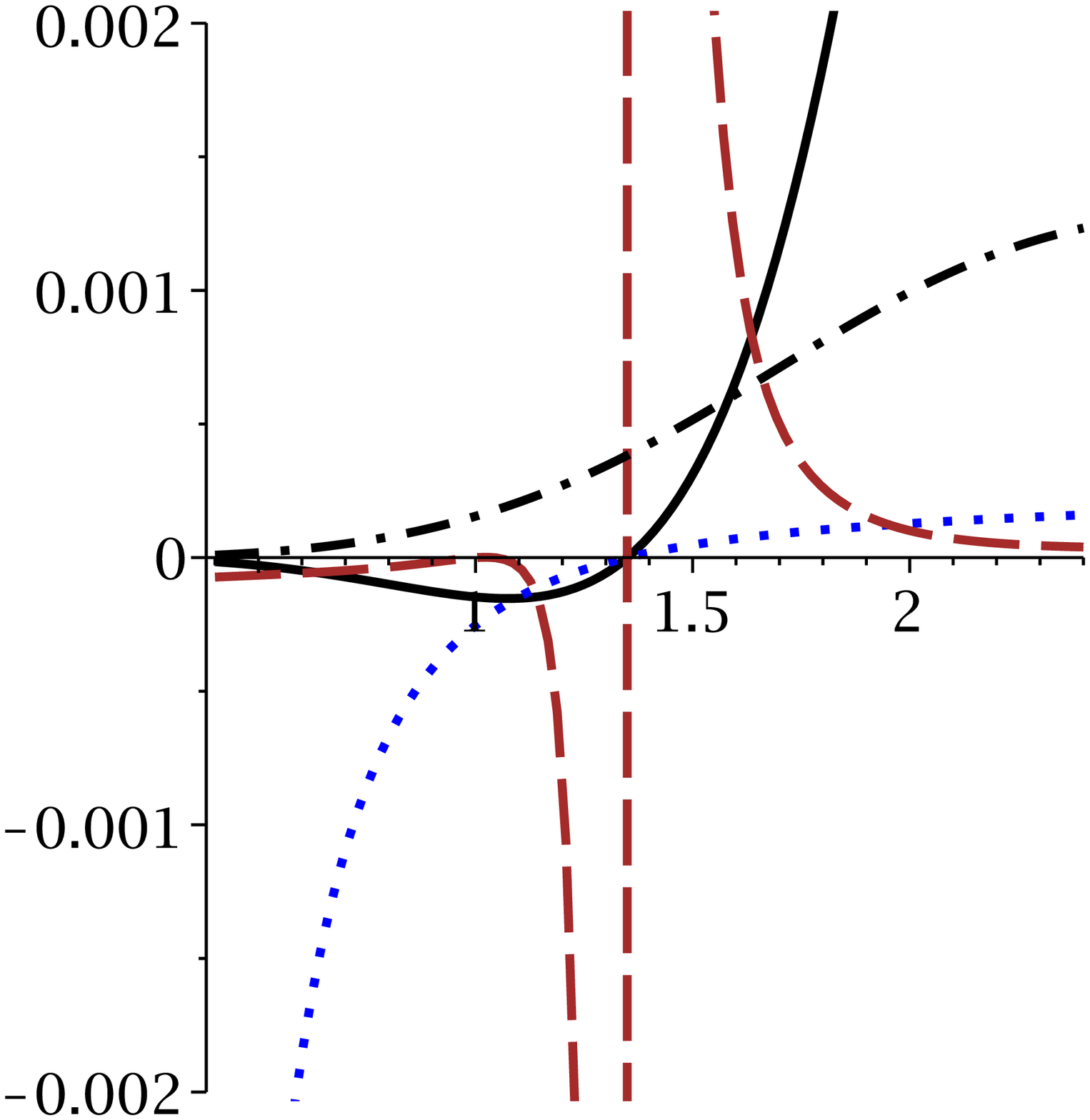} & \epsfxsize=6cm \epsffile{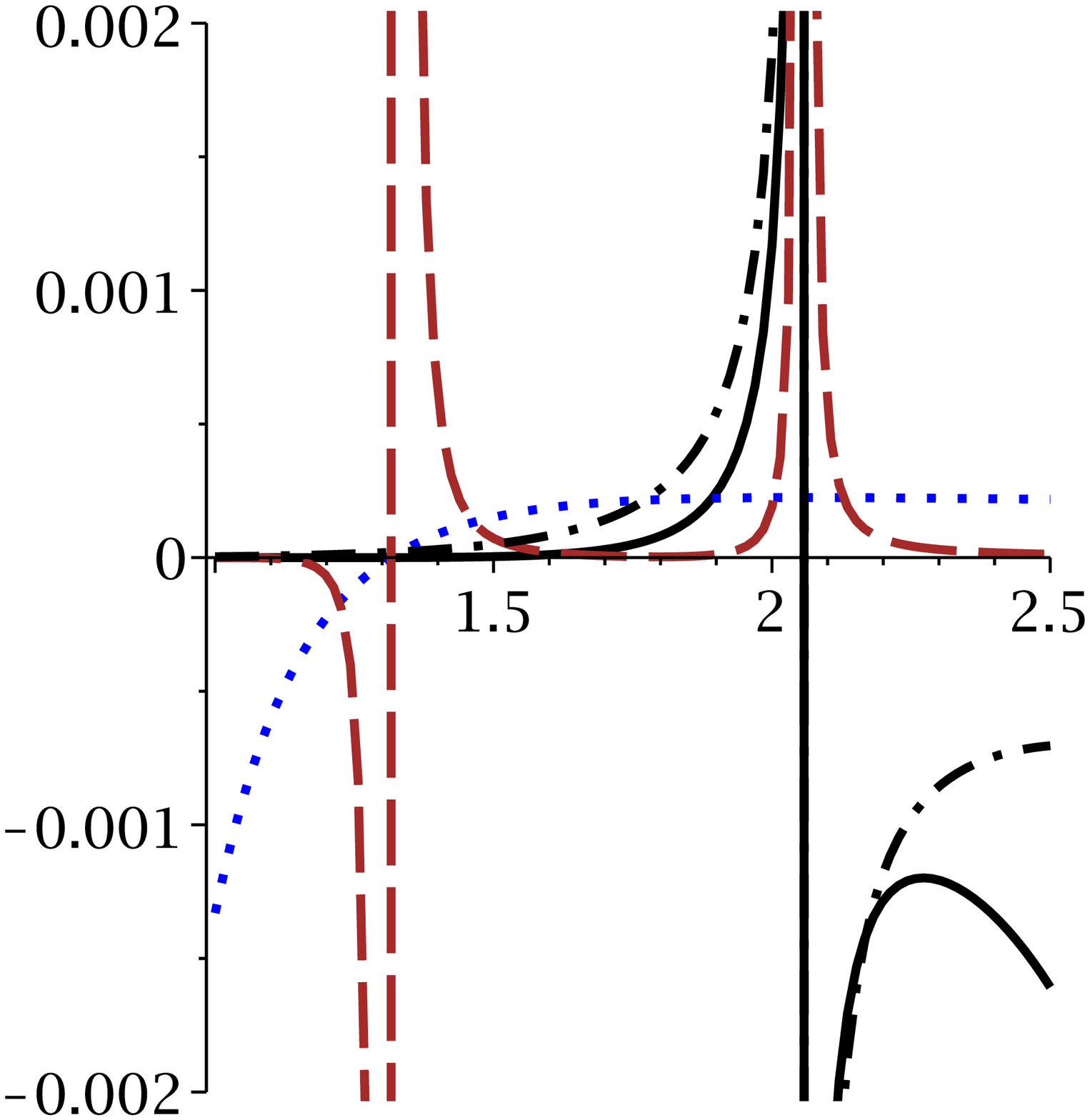} & %
\epsfxsize=6cm \epsffile{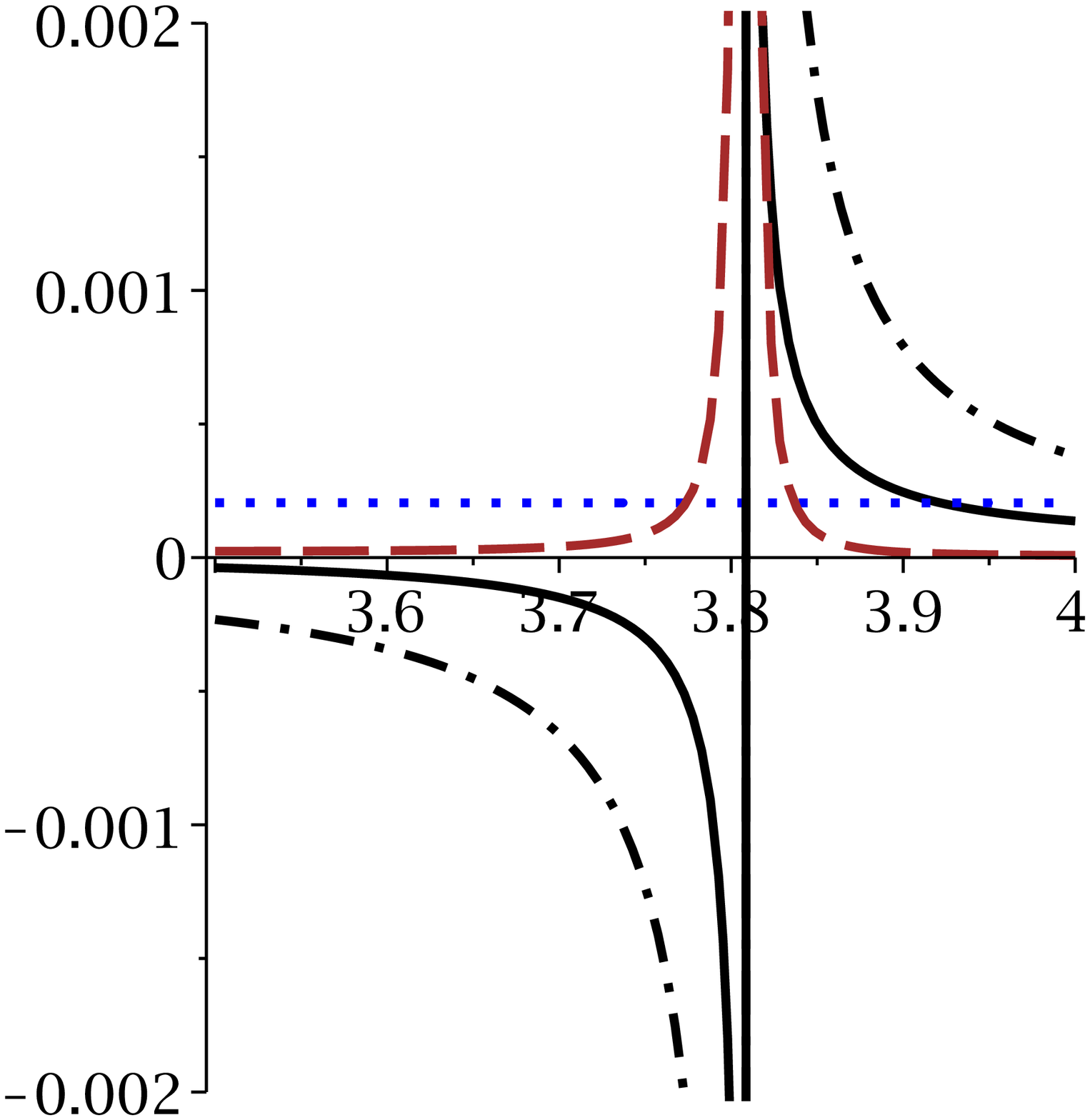}%
\end{array}
$%
\caption{\textbf{HPEM metric:} $\mathcal{R}$ (dashed line), T (dotted line),
$\protect\alpha$ (dashed-dotted line) and $C_{Q}$ (continues line) versus $%
r_{+}$ for $\Lambda=-1$, $q=1.1$ and $l=1$; Left panel: $n=4$, middle and
right panels: $n=7$ (\emph{For different scales}).}
\label{Fig5}
\end{figure}
\begin{figure}[tbp]
$%
\begin{array}{ccc}
\epsfxsize=6cm \epsffile{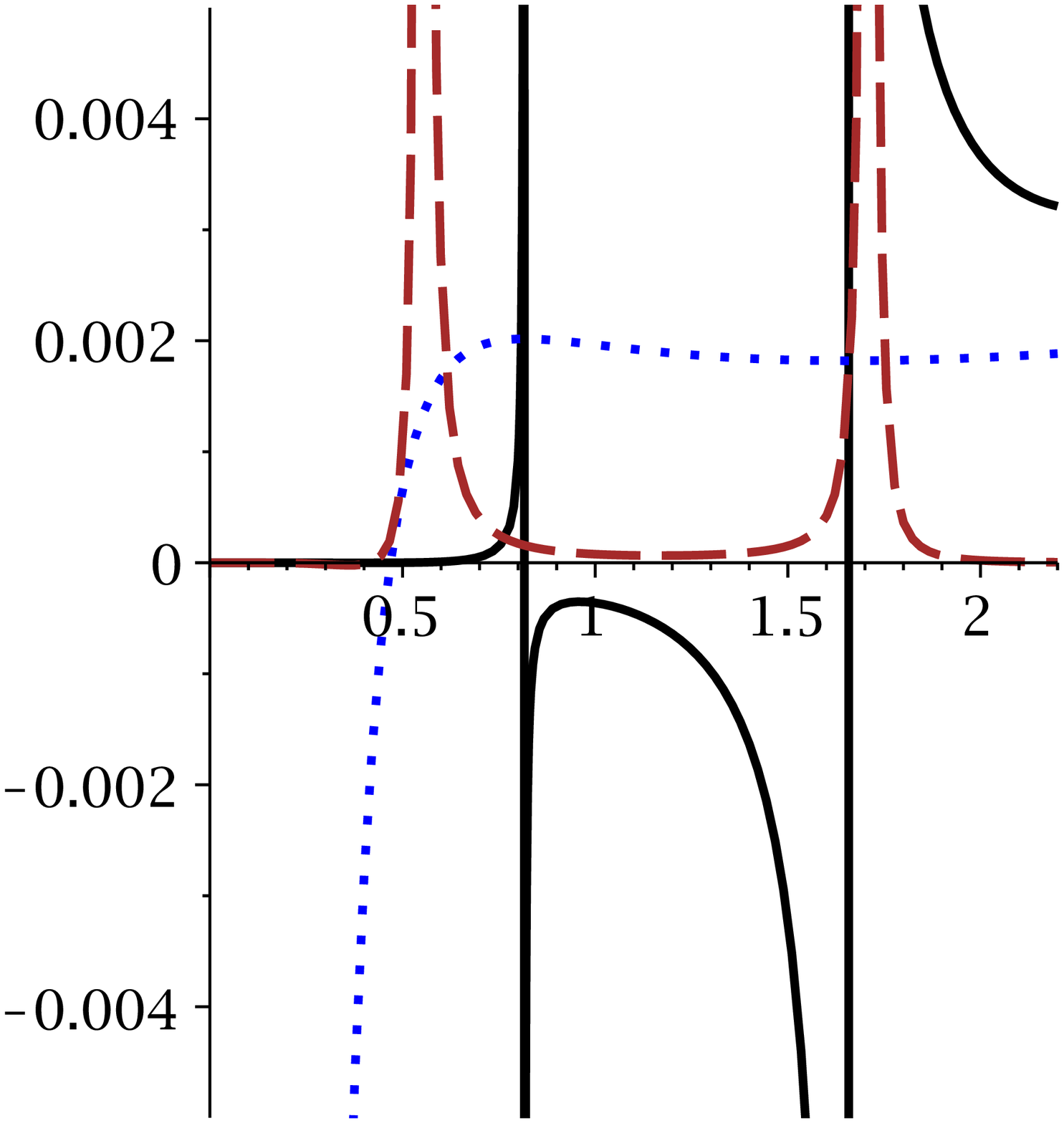} & \epsfxsize=6cm %
\epsffile{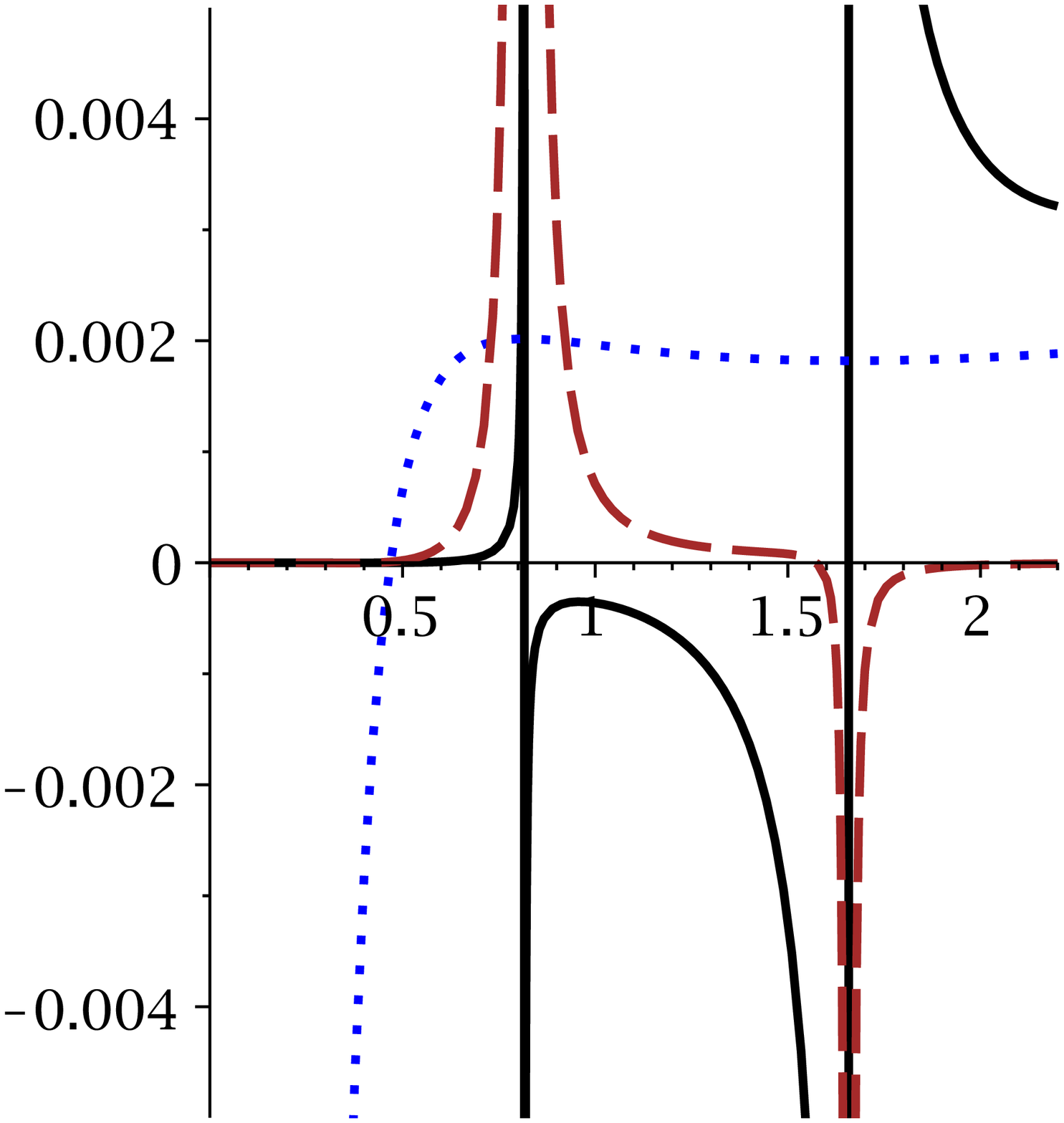} &
\end{array}
$%
\caption{$\mathcal{R}$ (dashed line), T (dotted line) and $C_{Q}$ (continues
line) versus $r_{+}$ for $\Lambda=-1$, $s=1.2$, $q=0.5$ and $n=4$; Left:
Weinhold metric, right panel: Quevedo metric (\emph{For different scales}).}
\label{Fig111}
\end{figure}

Depending on the choices of different parameters, for AdS space-time, the
temperature could be only an increasing function of the horizon radius with
one root (right panel of Fig. \ref{Fig1} and left panels of Figs. \ref{Fig2}
and \ref{Fig3}) or it may acquire extrema with one root (left and middle
panels of Fig. \ref{Fig1}; middle and right panels of Figs. \ref{Fig2} and %
\ref{Fig3}). The temperature and heat capacity share same root. This is the
bound point. For $r_{+}<r_{0}$ ($r_{0}$ is the root), solutions have
negative temperature and heat capacity. Therefore, in this region, solutions
are non-physical ones.

For small values of $q$, the temperature will have two extrema. In
places of these extrema, heat capacity is divergent (left and
middle panels of Fig. \ref{Fig1}). In other words, in places of
these extrema, system goes under second order phase transitions.
By increasing electric charge, these extrema are vanished and only
a bound point will remain (right panel of Fig. \ref{Fig1}). As for
nonlinearity parameter, its effects are opposite of those of
electric charge. In other words, by increasing nonlinearity
parameter, the extrema for temperature and divergencies for heat
capacity are formed (middle and right panels of Fig. \ref{Fig2})
while for small values of this parameter, only bound point is
observed (left panel of Fig. \ref{Fig2}). The effects of
dimensionality are similar to those observed for nonlinearity
parameter. Meaning that by increasing dimensionality, two extrema
are formed and system will acquire second order phase transitions
in its phase space in higher dimensions (Fig. \ref{Fig3}).

As for stability, if only bound point is observed, physical stable
solutions is observed only for $r_{0}<r_{+}$. The formation of
extrema modifies the stability conditions of these black holes. In
this case, bound point and two divergencies of the heat capacity
determine stable regions and phase transitions. Between bound
point and smaller divergence point, both temperature and heat
capacity are positive. Therefore, the solutions are physical and
stable in this region. On the other hand, between two
divergencies, temperature is positive while heat capacity is
negative, which leads to solutions being physical and unstable.
Finally after larger divergency, both heat capacity and
temperature are positive and solutions are stable. Thermodynamical
concepts indicate that system in unstable state goes under phase
transition and acquires stability. By taking this matter into
account, in smaller divergency a phase transition of larger to
smaller takes place while vice versa happens in larger divergency.

\subsubsection{CIM case}

First, lets study the conditions for the positivity/negativity of the
solutions. In order to have positive heat capacity, it needs to satisfy $%
A^{\prime }\times B^{\prime }>0$ in which%
\begin{eqnarray*}
A^{\prime } &=&2\Lambda r_{+}^{2n-1}+2^{\frac{n}{2}}\left( n-1\right)
q^{n}r_{+}^{n-1}-(n-1)(n-2)kr_{+}^{2n-3}, \\
&& \\
B^{\prime } &=&(n-2)kr_{+}^{n-2}+\frac{2\Lambda }{(n-1)}r_{+}^{n}-2^{\frac{n%
}{2}}\left( n-1\right) q^{n},
\end{eqnarray*}
where $A^{\prime}$ and $B^{\prime}$ are, respectively, the
numerator and the denominator of heat capacity.

It is not possible to obtain roots and divergencies of the heat
capacity for this case, analytically. Therefore, we employ
numerical methods and plot following diagrams for studying the
effects of different parameters (Figs. \ref{Fig4} and \ref{Fig5}).

It is evident that for small $q$, temperature has one root with
two extrema. The roots of temperature and heat capacity coincide
with each other while the extrema of temperature are matched with
divergencies of the heat capacity. These divergencies are phase
transition points. Increasing the electric charge leads to
elimination of phase transition points while there exists a root
for temperature and heat capacity (see Fig. \ref{Fig4} for more
details). The opposite behavior is observed for the effects of
dimensionality. Meaning that by increasing dimensionality, system
with one bound point, acquires one root and two divergencies. In
other words, by increasing dimensions, the thermodynamical
structure of the black holes is modified and black holes will
enjoy second order phase transition in their phase diagrams (Fig.
\ref{Fig5}).

\subsection{Geometrical thermodynamics}

Regarding ordinary laboratory systems (short-range interaction
property; entropy is proportional to volume), it is expected to
obtain the same thermodynamic properties for all ensembles.
However, taking into account the holography conception, one finds
that black holes are not ordinary systems (long-range interaction
property; entropy is proportional to area)
\cite{entropy1,entropy2,entropy3,entropy4,entropy5,entropy6,entropy7}.
Therefore, one may obtain different results in the context of
various ensembles of black hole thermodynamics \cite{QuevedoIV}.
Regardless of tracing ensemble dependency, it was shown that there
are some extra poles (or mismatch with other methods) in the Ricci
scalar of Weinhold, Ruppeiner and Quevedo metrics. Existence of
these anomalies/mismatches motivate one to build another Legendre
invariant thermodynamic metric which can be conformally related to
Quevedo metric. HPEM metric is the same as Quevedo metric up to a
conformal factor. Since the differences in the conformal factor
comes from the mass and its derivatives, it is expected that HPEM
enjoys Legendre invariancy, but with a different Legendre
multiplier. It is also worthwhile to mention that the Legendre
invariance alone is not sufficient to guarantee a unique picture
of thermodynamical metrics in terms of their curvatures
\cite{Bravetti}. In other words, in addition to Legendre
invariancy, we need to demand curvature invariancy under a change
of representation. As a result, it will be worthwhile to consider
the investigation of both Legendre and curvature invariancies.

Here, we will employ the geometrical thermodynamics to study both
bound and phase transition points of these black holes. The
geometrical thermodynamics is based on constructing phase space by
using one of the thermodynamical quantities as thermodynamical
potential. The information regarding bound and phase transition
points are stored in Ricci scalar of this phase space. In other
words, the divergencies of this Ricci scalar present bound and
phase transition points. (For more details regarding the meaning
of curvature scalar in geometrical thermodynamics, we refer the
reader to Ref. \cite{RuppeinerBook}). Depending on choice of
thermodynamical potential, the components of phase space differ
(this is due to fact that each thermodynamical quantity has its
specific extensive parameters). There are
several well known methods for constructing phase space; Weinhold \cite%
{WeinholdI,WeinholdII}, Ruppeiner \cite{RuppeinerI,RuppeinerII},
Quevedo \cite{QuevedoI,QuevedoII} and HPEM
\cite{HPEMI,HPEMII,HPEMIII,HPEMIV}. In order to investigate
thermodynamical properties of the system, the successful method
has divergencies which exactly are matched with bound and phase
transition points. The mentioned well-known thermodynamical
metrics in the context of geometrical thermodynamics are given by
\begin{equation}
ds^{2}=\left\{
\begin{array}{cc}
Mg_{ab}^{W}dX^{a}dX^{b} & Weinhold \\
&  \\
-T^{-1}Mg_{ab}^{R}dX^{a}dX^{b} & Ruppeiner \\
&  \\
\left( SM_{S}+QM_{Q}\right) \left( -M_{SS}dS^{2}+M_{QQ}dQ^{2}\right)  &
Quevedo \\
&  \\
S\frac{M_{S}}{M_{QQ}^{3}}\left( -M_{SS}dS^{2}+M_{QQ}dQ^{2}\right)  & HPEM%
\end{array}%
\right. ,  \label{metrics}
\end{equation}%
in which $M_{X}=\partial M/\partial X$ and $M_{XX}=\partial ^{2}M/\partial
X^{2}$. The denominators of their Ricci scalars will be \cite{HPEMI}
\begin{equation}
denom(\mathcal{R})=\left\{
\begin{array}{cc}
M^{2}\left( M_{SS}M_{QQ}-M_{SQ}^{2}\right) ^{2} & Weinhold \\
&  \\
M^{2}T\left( M_{SS}M_{QQ}-M_{SQ}^{2}\right) ^{2} & Ruppeiner \\
&  \\
M_{SS}^{2}M_{QQ}^{2}\left( SM_{S}+QM_{Q}\right) ^{3} & Quevedo \\
&  \\
2S^{3}M_{SS}^{2}M_{S}^{3} & HPEM%
\end{array}%
\right. .
\end{equation}

It is worthwhile to mention that Weinhold and Ruppeiner metrics
were formally introduced as the Hessian of the internal energy
(mass) and entropy, respectively
\begin{equation}
ds^{2}=\left\{
\begin{array}{cc}
\frac{\partial^{2} M}{\partial X^{a} \partial X^{b}}dX^{a}dX^{b} & Weinhold \\
&  \\
-\frac{\partial^{2} S}{\partial X^{a} \partial X^{b}}dX^{a}dX^{b}
& Ruppeiner
\end{array}%
\right. .  \label{met2new}
\end{equation}%

Later, it was shown that these two methods are conformally related
by adding terms which could be seen in Eq. (\ref{metrics}). In
this paper, we employ the forms which are given in Eq.
(\ref{metrics}) for Weinhold and Ruppeiner methods.

Before we continue our study, it is worthwhile to point out a few
remarks regarding HPEM metric. The HPEM and Quevedo metrics have
similar structure up to a conformal factor with same signature
which is ($-+++...$). In Ref. \cite{QuevedoIII}, it was shown that
it is possible to obtain an infinite number of Legendre invariant
metrics. The simplest way to construct Legendre invariant metrics
is to apply a conformal transformation. Considering the similar
structure of HPEM and Quevedo metrics with same signature, it is
expected that with a different Legendre multiplier, HPEM also
enjoys Legendre invariancy. The studies conducted in the context
of HPEM metric showed that signature of the curvature scalar of
this metric around divergencies provides the possibility to
recognizing the type of phase transition (smaller to larger or
vice versa). In addition, by studying signature, it is possible to
differ bound and phase transition points from one another. These
differences in signature around divergencies also enables one to
have a picture regarding stability/instability of black holes. One
of the main properties of this metric is to include bound points
in its divergencies. This results into having a better and more
complete picture regarding thermodynamical structure of the black
holes comparing to previous metrics. In addition, it was shown
that the results of this metric is in agreement with other methods
of studying thermodynamical structure of the black holes such as
the heat capacity, van der Waals like diagrams and etc
\cite{int,ourpapers1,ourpapers3}.

Existence of $M_{S}$ and $M_{SS}$ in denominator of the HPEM metric ensures
that bound and phase transition points are matched with divergencies of the
Ricci scalar without any extra divergency. As for Quevedo metric, since $%
M_{SS}$ is present in denominator of this metric, phase transition points
are matched with divergencies of the Ricci scalar. On the other hand, the
bound points are matched only when
\begin{equation}
M_{S}=-\frac{QM_{Q}}{S},  \label{cond}
\end{equation}%
which in general is not satisfied. In addition, the presence of $%
M_{QQ}$ may lead to existence of extra divergencies which are not
matched with bound and phase transition points. Such a case has
been reported for several black holes before
\cite{HPEMI,HPEMII,HPEMIII,HPEMIV}. Analytical evaluation shows
that $M_{QQ}$ for these black holes will not lead to any extra
divergency, but $M_{S}=-\frac{QM_{Q}}{S}$ is not satisfied.

Recently, in Ref. \cite{newQ}, it was pointed out that Quevedo
metric should have the following form instead of the one pointed
out in Eq. (\ref{metrics}),
\begin{equation}
ds^{2}=M\left( -M_{SS}dS^{2}+M_{QQ}dQ^{2}\right) ,
\end{equation}%
where the denominator of its Ricci scalar will be
\begin{equation}
denom(\mathcal{R})=M^{3}M_{SS}^{2}M_{QQ}^{2}.
\end{equation}

Here, the condition which was observed for the previous case
(\ref{cond}) is eliminated for the Ricci scalar. The divergencies
of heat capacity are included through the presence of $M_{SS}^{2}$
term. Meanwhile, the existence of $M_{QQ}$ results into presence
of divergencies for Ricci scalar which are not consistent with any
phase transition point. Recently, through several studies it was
shown that it is possible for the mass of the black holes to have
root(s) \cite{roots1,roots2}. This indicates that for this form of
the Quevedo metric, the existence of $M$ and its roots may provide
the possibility of the existence of divergencies in its Ricci
scalar which are not matched with any phase transition point.
Therefore, cases of mismatch between divergencies of the Ricci
scalar of this Quevedo metric and phase transition points are
possible. It is worthwhile to mention that this case of Quevedo
metric does not include bound points in divergencies of its Ricci
scalar.

In case of Weinhold and Ruppeiner, it was not analytically possible to see
whether divergencies of Ricci scalars of these methods coincide completely
with bound and phase transition points. Therefore, we employed numerical
methods and plotted some diagrams (see Fig. \ref{Fig111}).

Evidently, employing Weinhold and Quevedo metrics will lead to
inconsistent results regarding bound and phase transition points.
In other words, by taking a closer look at the plotted diagrams
(see Fig. \ref{Fig111}), it is obvious that there is a mismatch
between divergencies of Ricci scalar of Weinhold metric and bound
and phase transition points. In addition, the Quevedo metrics
fails to produce divergency in its Ricci scalar which is
coincidence with bound point. On the contrary, divergencies of the
Ricci scalar of HPEM metric coincide with the bound and phase
transition points completely (see Figs. \ref{Fig1}-\ref{Fig5}).

One of the interesting properties of HPEM metric is the behavior
of its Ricci scalar around bound and phase transition points. If
the divergency is related to the phase transition of larger to
smaller black holes, the signature of Ricci scalar around it, is
positive. On the contrary, for phase transition of smaller to
larger black holes, the sign of Ricci scalar around this
divergence point is negative. As for the bound point,
interestingly, the signature of Ricci scalar changes around this
divergence point. These changes in the signature of Ricci scalar
enable one to recognize the type of points without prior knowledge
about heat capacity. As for Ruppeiner, since it has conformal
relation with Weinhold metric, same results are applicable for
phase transition points. But, due to the existence of temperature
in denominator of Ricci scalar of Ruppeiner, the bound points are
matched with some of the divergencies.

To summarize, we saw that employing Weinhold, Ruppeiner and
Quevedo metrics, leads to the existence of divergencies in the
Ricci scalar which are not match with the obtained bound and phase
transitions of heat capacity, while the HPEM metric provides a
consistent number of divergencies in the places of bound and phase
transition points. Furthermore, the characteristic behavior of
Ricci scalar of HPEM metric enables one to recognize the type of
point and phase transition.

\subsection{Critical behavior in extended phase space}

In this section, we will conduct a study regarding possible
existence of critical and van der Waals like behavior for the
solutions. Recently, there has been a novel interpretation for the
cosmological constant in AdS black holes. Instead of considering
this parameter as a fixed one, it was proposed to consider it as a
thermodynamical variable. By doing so, the thermodynamical
structure of black holes is enriched and it is possible to
introduce new phenomenologies such as van der Waals like behavior
and phase transition for black holes. The van der Waals like
behavior of these black holes through usual method was
investigated in Ref. \cite{CosmP9}. Here, we would like to study
critical values through a new method which was introduced in Ref.
\cite{int}. In this method, the cosmological constant is replaced
by its proportional pressure in the heat capacity. Then, by using
denominator of the heat capacity and solving it with respect to
the pressure, a relation for pressure is obtained which is
different from the usual equation of state. The maximum of this
relation marks a place in which phase transition takes place. In
other words, the maximum of this relation in $P-r_{+}$ diagram is
corresponding to the critical pressure. This method was employed
in several papers and it was shown to have consistent results with
those extracted through usual methods
\cite{int,ourpapers1,ourpapers3}.

The proportionality between cosmological constant and pressure is given by
\begin{equation}
P=-\frac{\Lambda }{8\pi }.  \label{PP}
\end{equation}

By replacing the cosmological constant with its corresponding
thermodynamical pressure, the mass of black holes will play the
role of enthalpy instead of internal energy. Using this new
concept for the mass of black holes, it is a matter of calculation
to obtain the volume of black holes, conjugate to the
thermodynamical pressure, in following form
\begin{equation}
V=\left( \frac{\partial H}{\partial P} \right)=\left(
\frac{\partial M}{\partial  P} \right)=\frac{r_{+}^{n-2}}{n-2}.
\end{equation}

Evidently, the volume is a smooth function of the horizon radius
and it is only horizon radius and dimension dependent. This
relation between the horizon radius and volume provides the
possibility of employing horizon radius instead of volume to study
phase diagrams of the black holes. In other words, the horizon
radii in which system goes under phase transition could be used to
obtain critical volume of our thermodynamical systems.

Now, by using the heat capacity (\ref{Heat}) with (\ref{PP}) and
solving the denominator of heat capacity with respect to
thermodynamical pressure, one obtains the following relation for
pressure
\begin{equation}
P=\left\{
\begin{array}{cc}
\begin{array}{c}
\frac{\left( n-1\right) }{16\pi r_{+}^{n}}\left[ \left( n-2\right)
kr_{+}^{n-2}-2^{\frac{n}{2}}\left( n-1\right) q^{n}\right] \\
\\
\end{array}
&
\begin{array}{c}
s=\frac{n}{2} \\
\\
\end{array}
\\
\frac{\left( 2\left( 2-n\right) s-1\right) \left[ \frac{(n-1)(n-2s)^{2}q^{2}%
}{(n-2)(2s-1)^{2}}\right] ^{s}r_{+}^{\frac{6s}{2s-1}}+k(n-1)(n-2)r_{+}^{%
\frac{2\left( ns+1\right) }{2s-1}}}{16\pi r_{+}^{\frac{2s(n+2)}{2s-1}}} &
otherwise%
\end{array}%
\right. .  \label{P}
\end{equation}

Using this relation, we are in a position to study possible
existence of phase transition for these black holes. Obtained
relation for pressure consists two terms; charge term and
topological term. The charge term has always negative contribution
to pressure. Therefore, the existence of critical behavior is
determined by the topological term. That being said, one can
automatically conclude that for $k=0,-1$, no critical behavior
exists. In other words, for these two cases, the pressure will be
negative which indicates that no critical behavior is observed.
Regarding the fact that maximum of the pressure (\ref{P}) is where
second order phase transition takes place, one can, analytically,
show that the maximum of this pressure is located at the following
horizon radius
\begin{equation}
r_{critical/\max }=\left\{
\begin{array}{cc}
\begin{array}{c}
\left( \frac{2^{\frac{\left( n-2\right) }{2}}n\left( n-1\right) q^{n}}{k(n-2)%
}\right) ^{\frac{1}{n-2}} \\
\\
\end{array}
&
\begin{array}{c}
s=\frac{n}{2} \\
\\
\end{array}
\\
\left( \frac{2s^{2}(n-2)+s}{k(n-2)(2s-1)}\right) ^{\frac{2s-1}{s(n-3)+1}%
}\left( \frac{(n-1)(n-2s)^{2}}{(n-2)(2s-1)^{2}}q^{2}\right) ^{\frac{s(2s-1)}{%
s(n-3)+1}} & otherwise%
\end{array}%
\right. ,
\end{equation}%
where by replacing the horizon radius, one finds
\begin{eqnarray}
&& \\
P_{critical/\max } &=&\left\{
\begin{array}{cc}
\begin{array}{c}
\frac{k\left( n-1\right) (n-2)^{2}}{16n\pi }\left( \frac{2^{\frac{\left(
n-2\right) }{2}}n\left( n-1\right) q^{n}}{k(n-2)}\right) ^{\frac{2}{2-n}} \\
\\
\end{array}
&
\begin{array}{c}
s=\frac{n}{2} \\
\\
\end{array}
\\
\frac{{k}^{{\frac{s(n+1)-1}{s(n-3)+1}}}{\Upsilon }}{16\pi (2s-1)^{{\frac{%
2-8s^{2}}{1+(n-3)s}}}}\left( 1-\frac{2k\left[ \frac{(n-2)(2s-1)^{2}}{(n-1)}%
\right] ^{{\frac{(n-3)s^{2}+(n-2)s+1}{s(n-3)+1}}}}{s^{2}\left[ 2s(n-2)+1%
\right] \left[ q^{2}(n-2s)^{2}\right] ^{s}}\right) & otherwise%
\end{array}%
\right. ,
\end{eqnarray}%
in which
\begin{equation*}
{\Upsilon }=\frac{(n-2)^{{\frac{4s^{2}+s(n-1)-1}{1+(n-3)s}}}(n-1)^{{\frac{%
1-4s^{2}+s(n-1)}{1+(n-3)s}}}}{\left\{ q^{2s}(n-2s)^{2s}\left[ 1+2s(n-2)%
\right] s\right\} ^{{\frac{2(2s-1)}{1+\left( n-3\right) s}}}}.
\end{equation*}

\begin{figure}[tbp]
$%
\begin{array}{cc}
\epsfxsize=6cm \epsffile{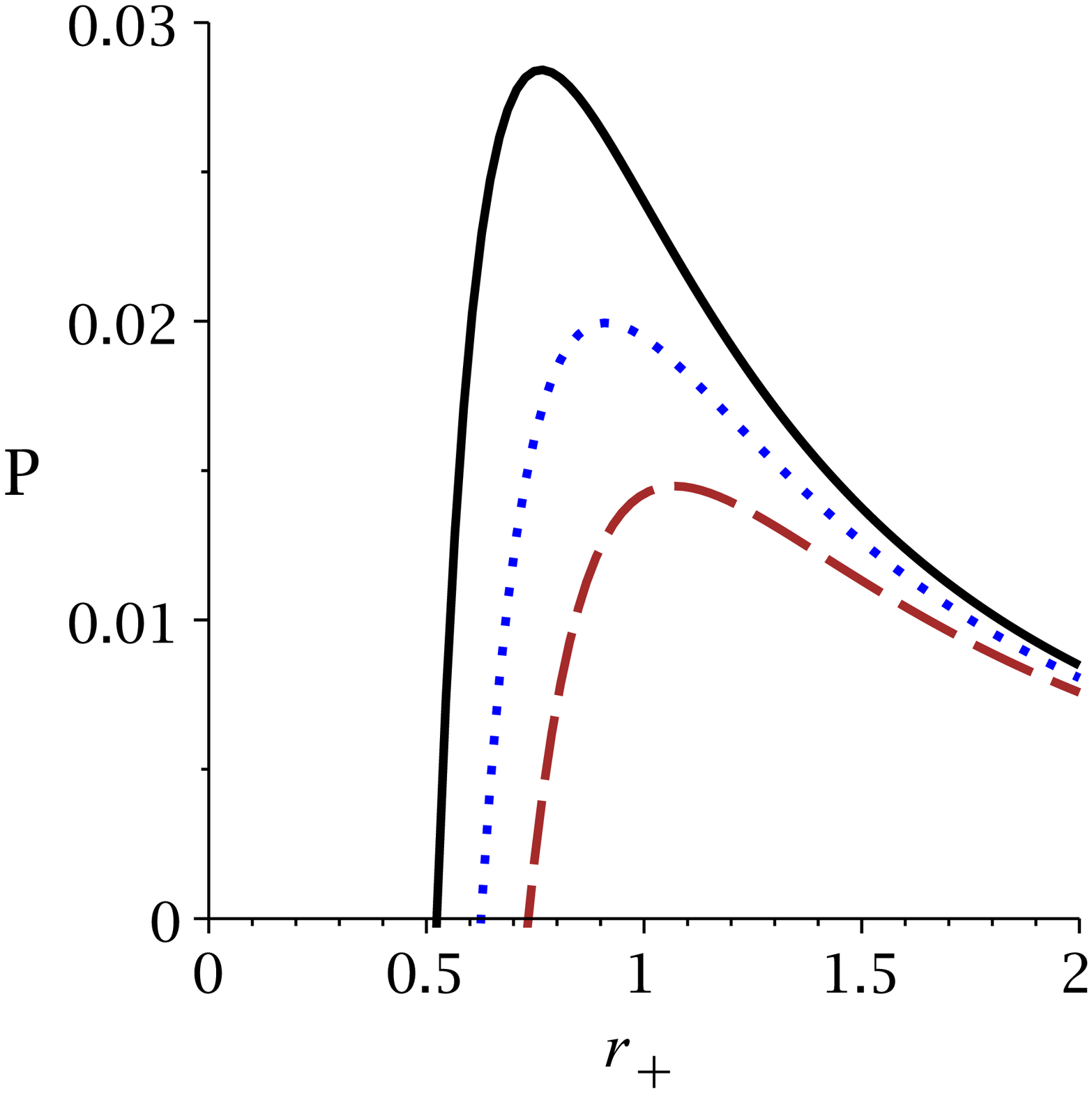} & \epsfxsize=6cm \epsffile{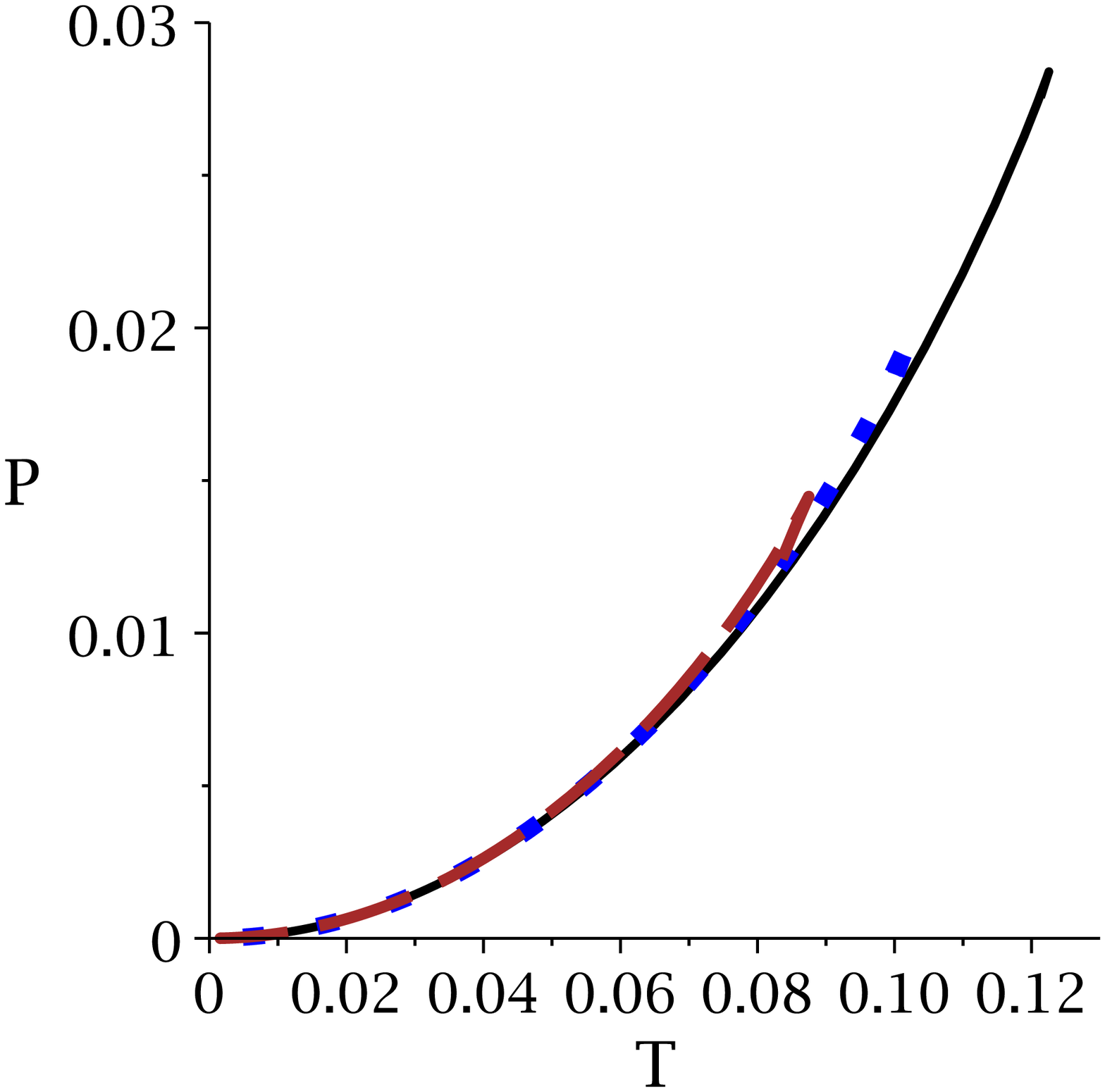}%
\end{array}
$%
\caption{Left panel: $P$ versus $r_{+}$ diagrams; Right panel: $P$ versus $T$
diagrams. \newline
for $k=1$, $n=3$ and $s=1.2$; $q=0.9$ (continues line), $q=1$ (dashed line)
and $q=1.1$ (dotted line). }
\label{Fig6}
\end{figure}

\begin{figure}[tbp]
$%
\begin{array}{cc}
\epsfxsize=6cm \epsffile{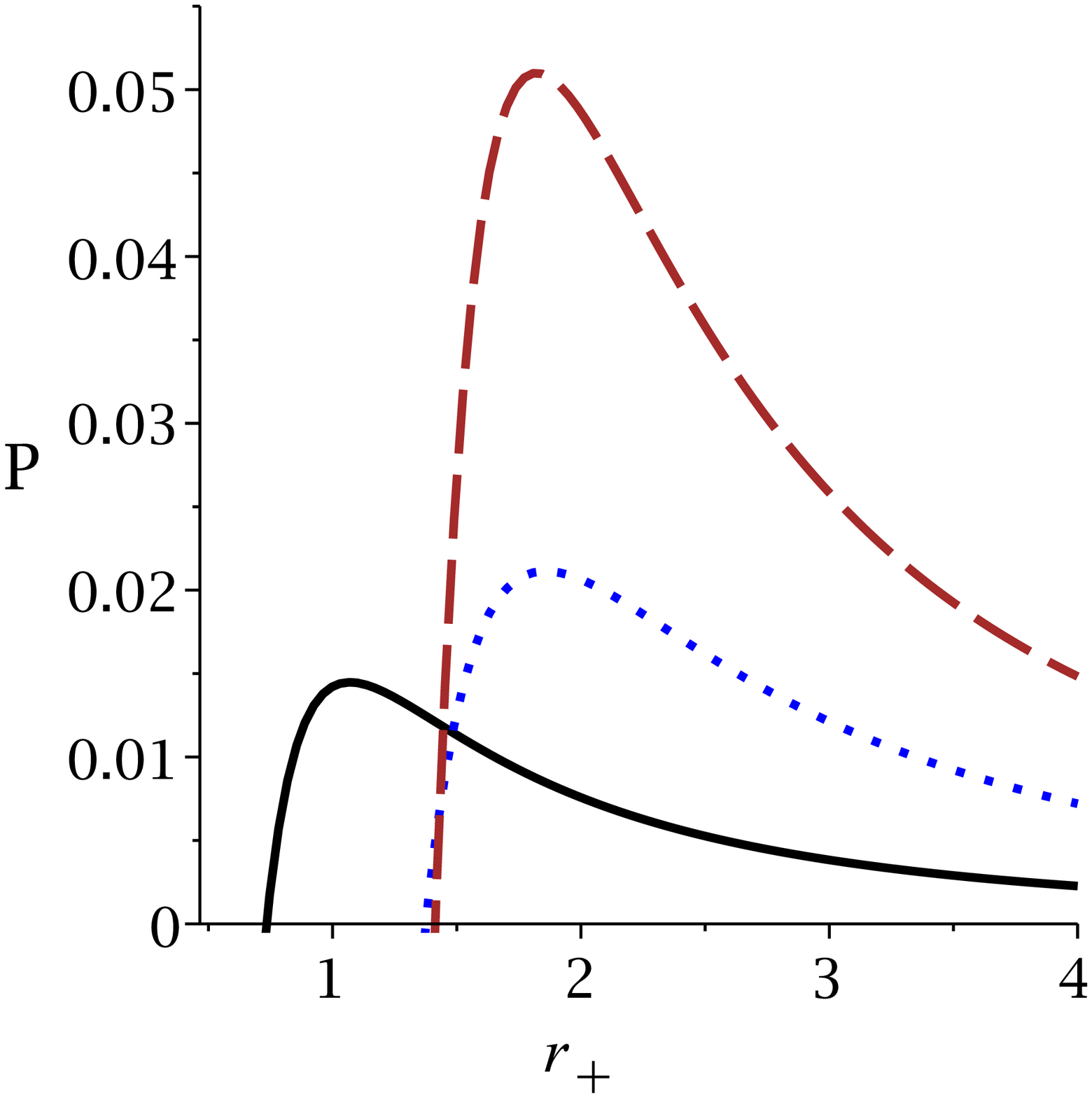} & \epsfxsize=6cm \epsffile{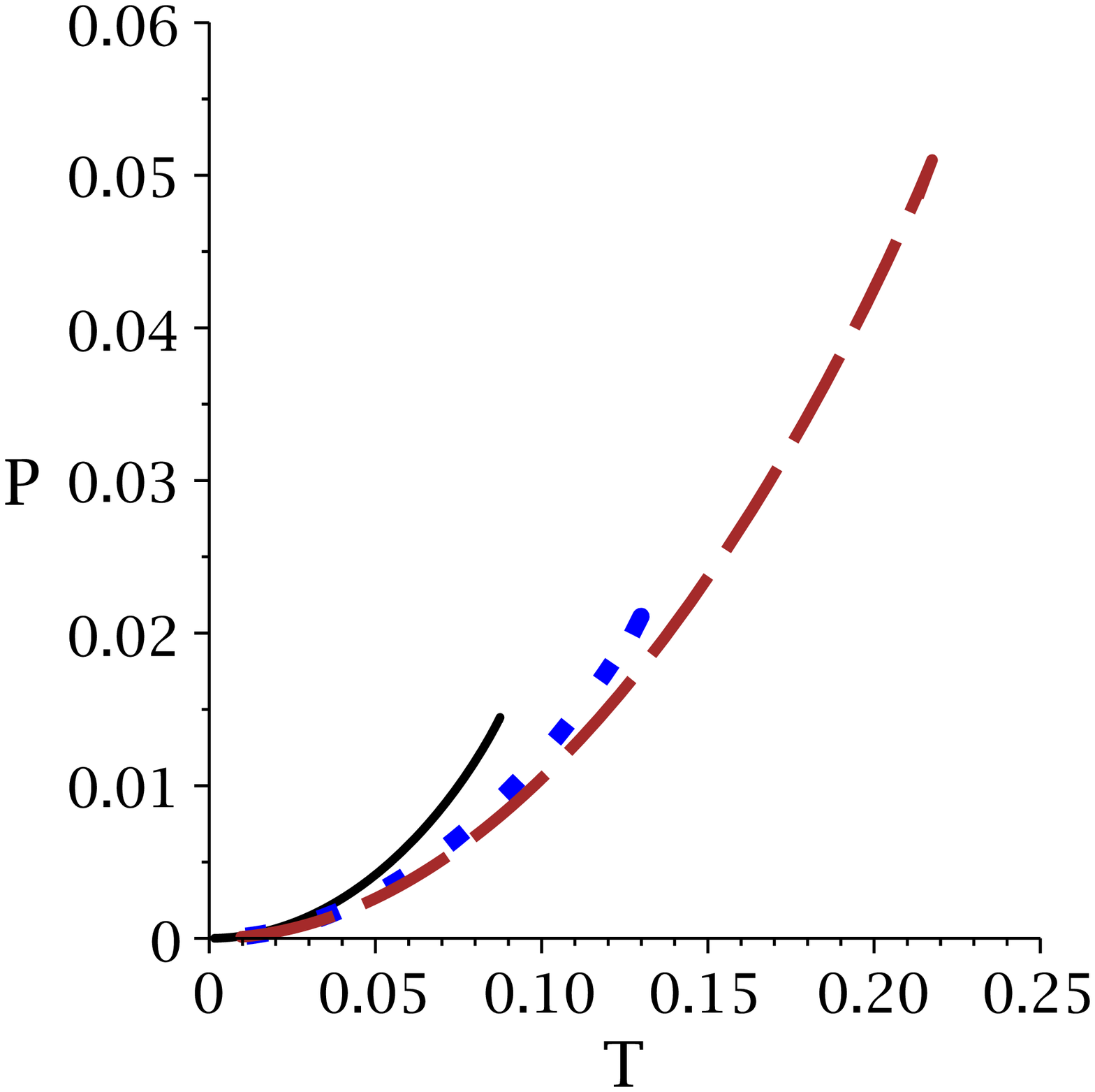}%
\end{array}
$%
\caption{Left panel: $P$ versus $r_{+}$ diagrams; Right panel: $P$ versus $T$
diagrams. \newline
for $k=1$, $s=1.2$ and $q=1.1$; $n=3$ (continues line), $n=4$ (dashed line)
and $n=5$ (dotted line). }
\label{Fig7}
\end{figure}

\begin{figure}[tbp]
$%
\begin{array}{cc}
\epsfxsize=6cm \epsffile{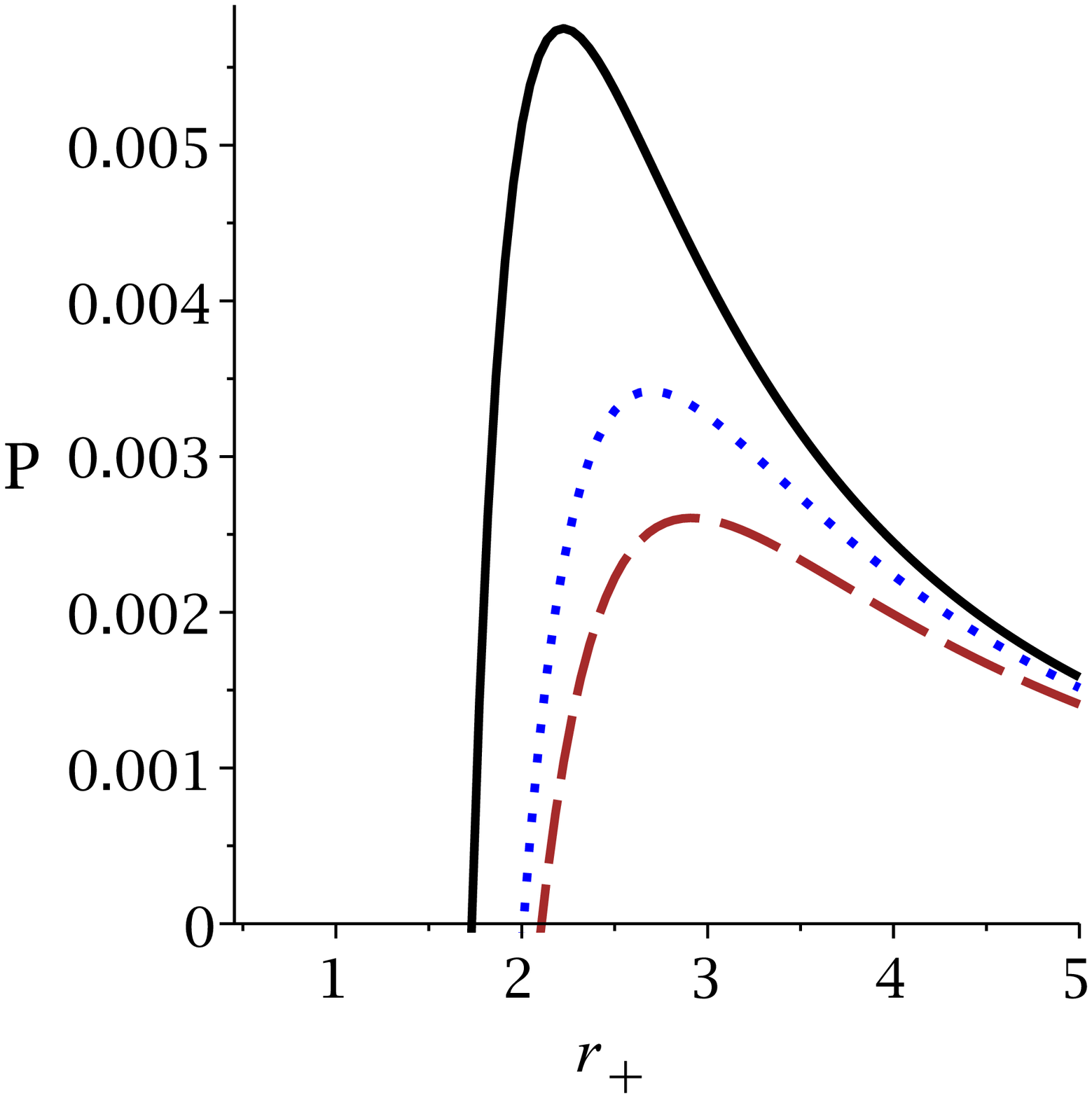} & \epsfxsize=6cm \epsffile{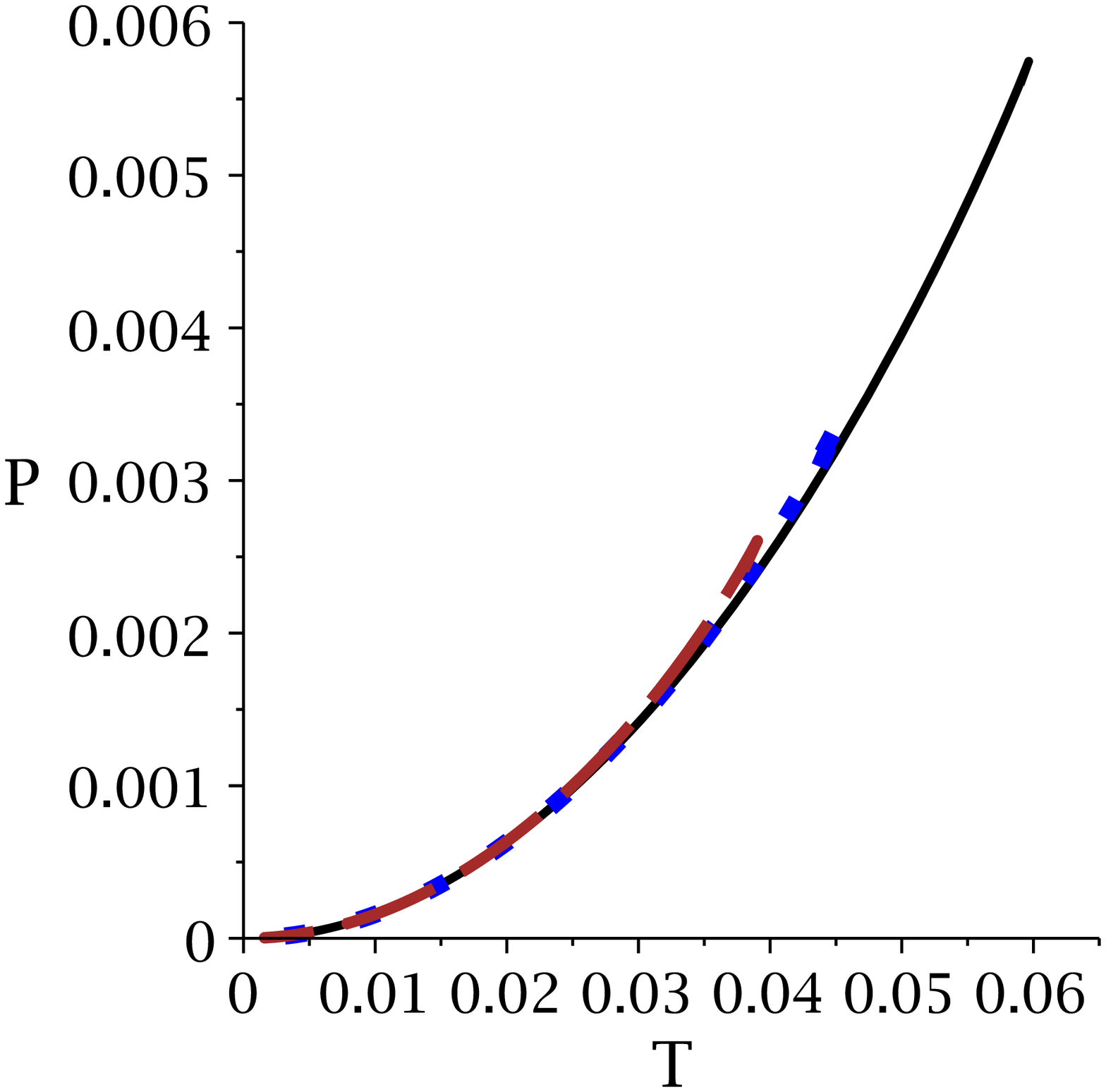}%
\end{array}
$%
\caption{Left panel: $P$ versus $r_{+}$ diagrams; Right panel: $P$ versus $T$
diagrams. \newline
for $k=1$, $n=3$ and $q=1.1$; $s=0.7$ (continues line), $s=0.8$ (dashed
line) and $s=0.9$ (dotted line). }
\label{Fig8}
\end{figure}

\begin{figure}[tbp]
$%
\begin{array}{cc}
\epsfxsize=6cm \epsffile{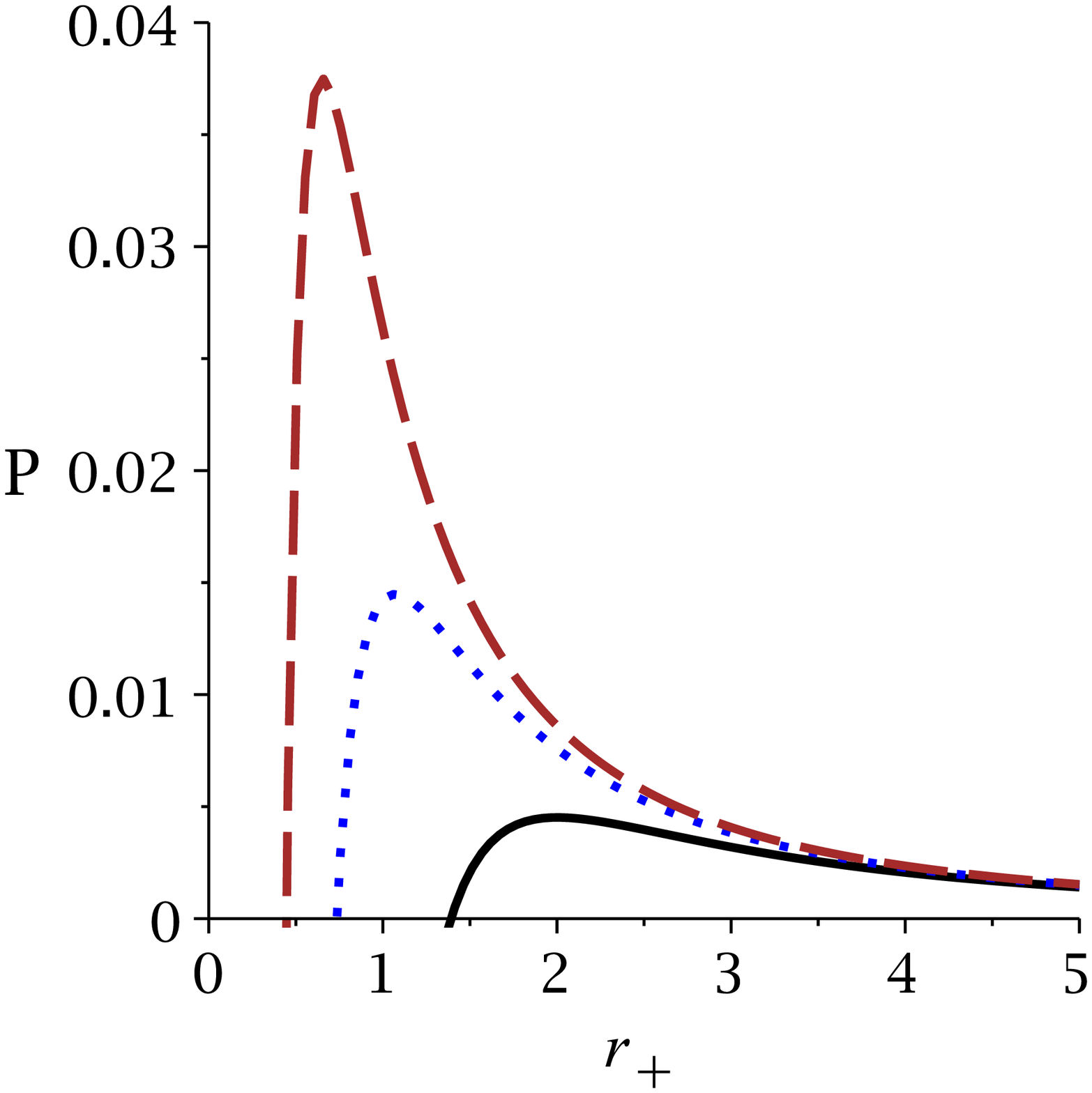} & \epsfxsize=6cm \epsffile{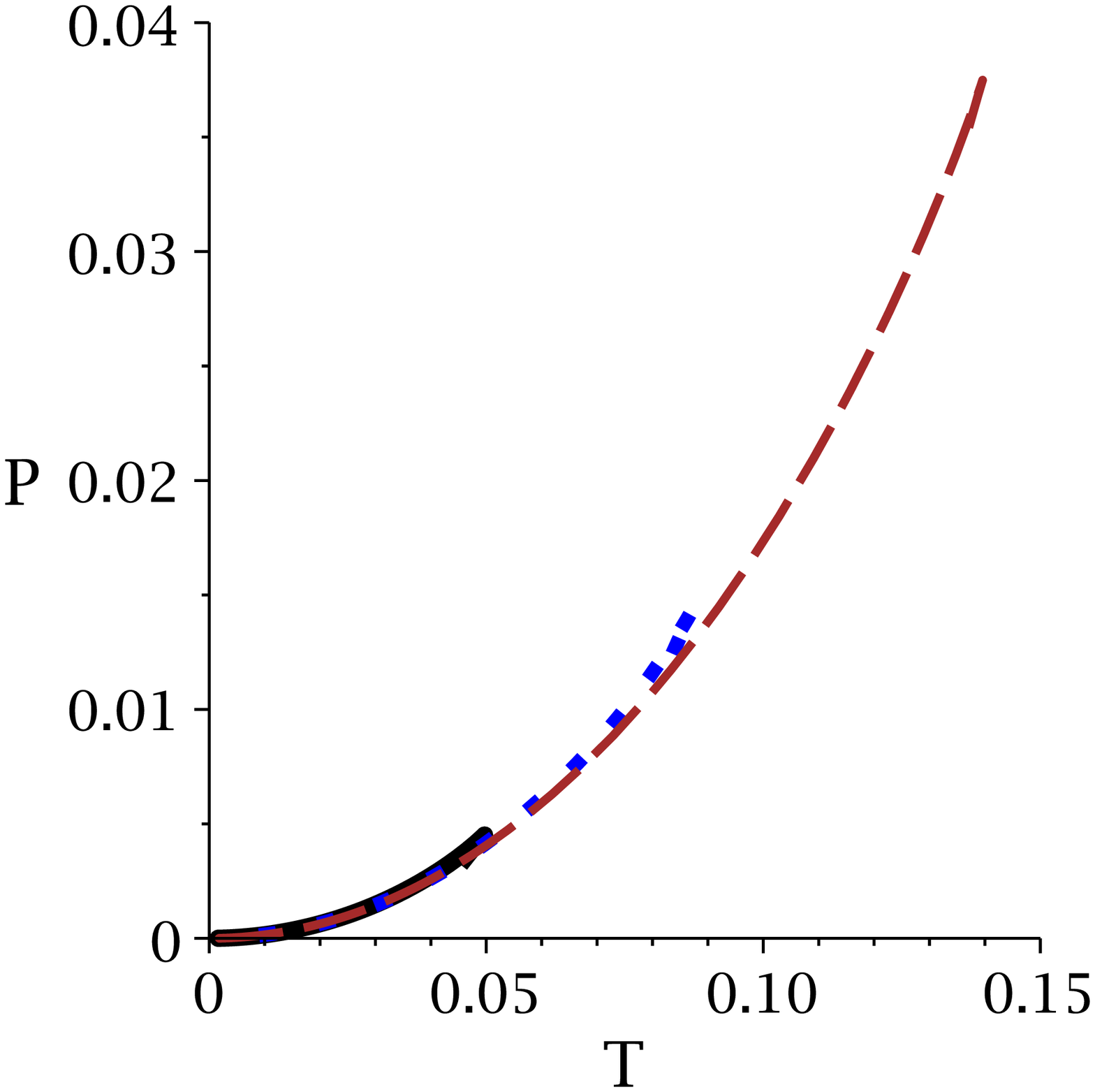}%
\end{array}
$%
\caption{Left panel: $P$ versus $r_{+}$ diagrams; Right panel: $P$ versus $T$
diagrams. \newline
for $k=1$, $n=3$ and $q=1.1$; $s=1.1$ (continues line), $s=1.2$ (dashed
line) and $s=1.25$ (dotted line). }
\label{Fig9}
\end{figure}

Obtained relation for the critical pressure indicates that the
positivity of critical pressure is subject to variation of
different parameters. It is evident that for Ricci-flat horizon
($k=0$), there is no positive critical pressure, whereas for
$k=-1$, one can obtain a real positive critical pressure when
$\frac{s(n+1)-1}{s(n-3)+1}\in \mathbb{N}$. In the case of
spherical horizon ($k=1$), the sign of $P_{critical/\max}$ is
depending on the values of nonlinearity, dimensionality and
electric charge. One may find that for sufficiently large values
of $q$, $s$ and $n$, the second term will be less than one and one
can obtain a positive critical pressure, while for usual values of
the mentioned parameter, $P_{critical/\max}$ can be negative and
one concludes there is no critical behavior. In order to show the
effects of different parameters on the critical behavior of
system, we plot various $P-r_{+}$ and $P-T$ diagrams (Figs.
\ref{Fig6}--\ref{Fig9}).

\begin{figure}[tbp]
$%
\begin{array}{cc}
\epsfxsize=6cm \epsffile{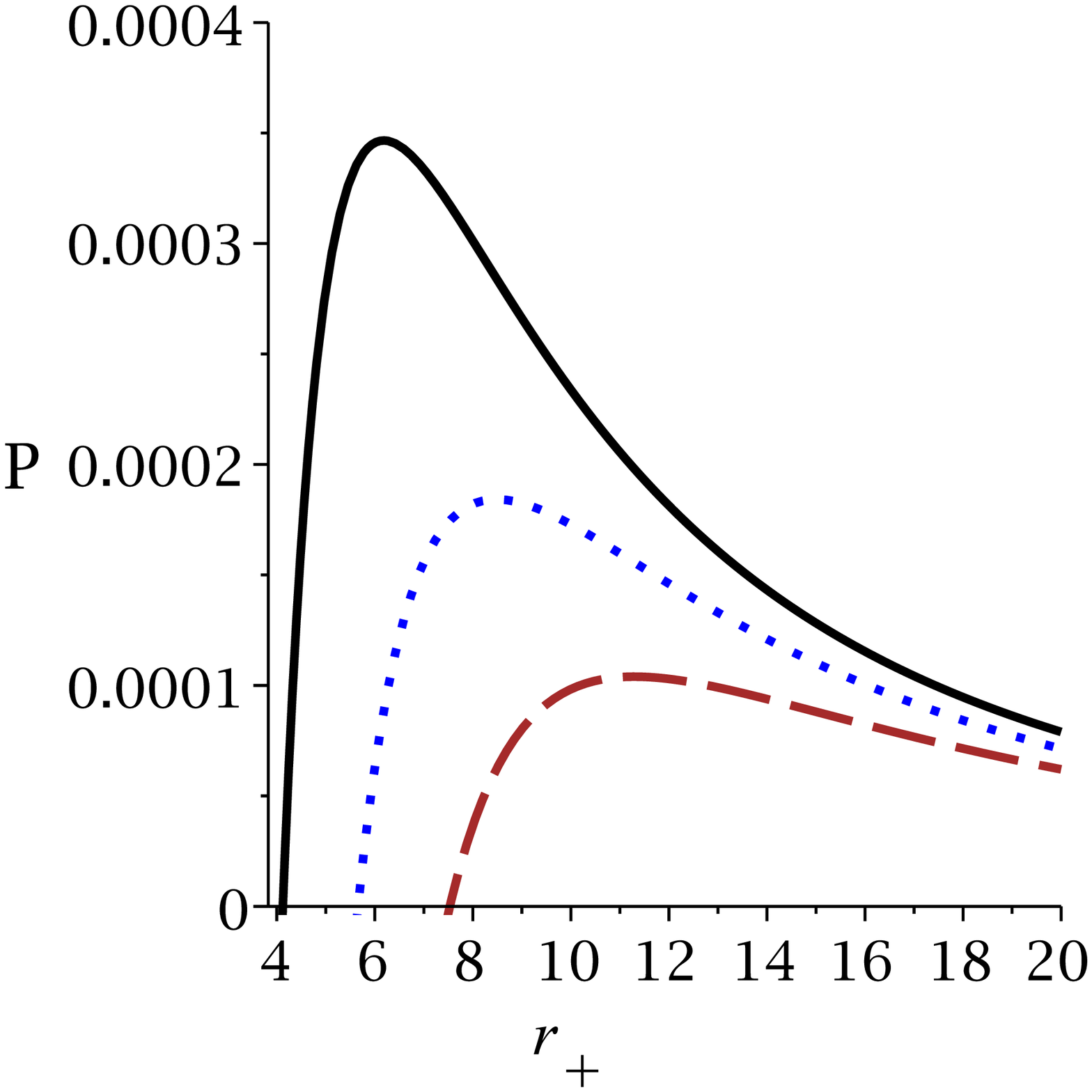} & \epsfxsize=6cm \epsffile{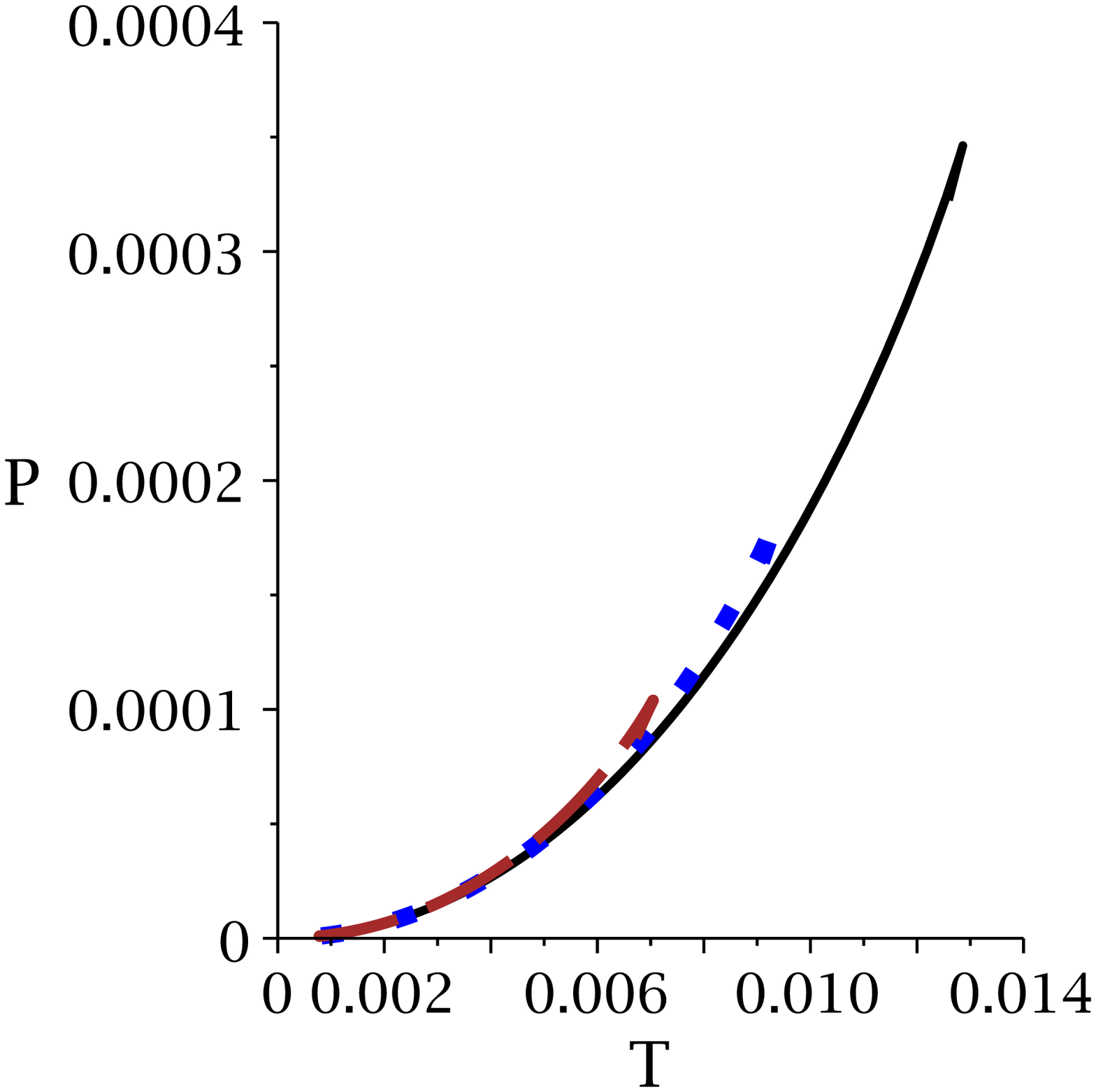}%
\end{array}
$%
\caption{Left panel: $P$ versus $r_{+}$ diagrams; Right panel: $P$ versus $T$
diagrams. \newline
for $k=1$, $n=3$ and $l=1$; $q=0.9$ (continues line), $q=1$ (dashed line)
and $q=1.1$ (dotted line). }
\label{Fig10}
\end{figure}
\begin{figure}[tbp]
$%
\begin{array}{cc}
\epsfxsize=6cm \epsffile{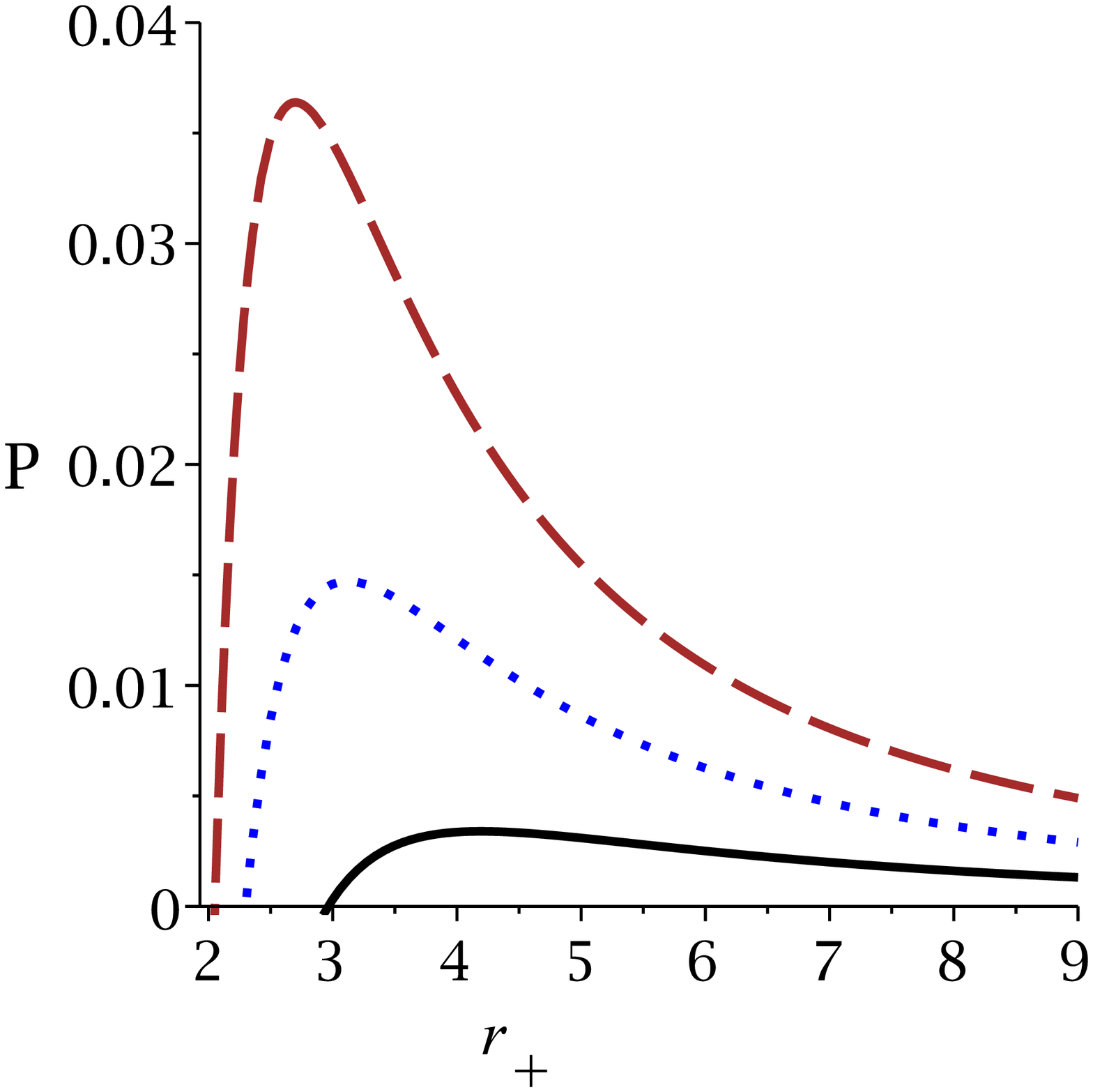} & \epsfxsize=6cm \epsffile{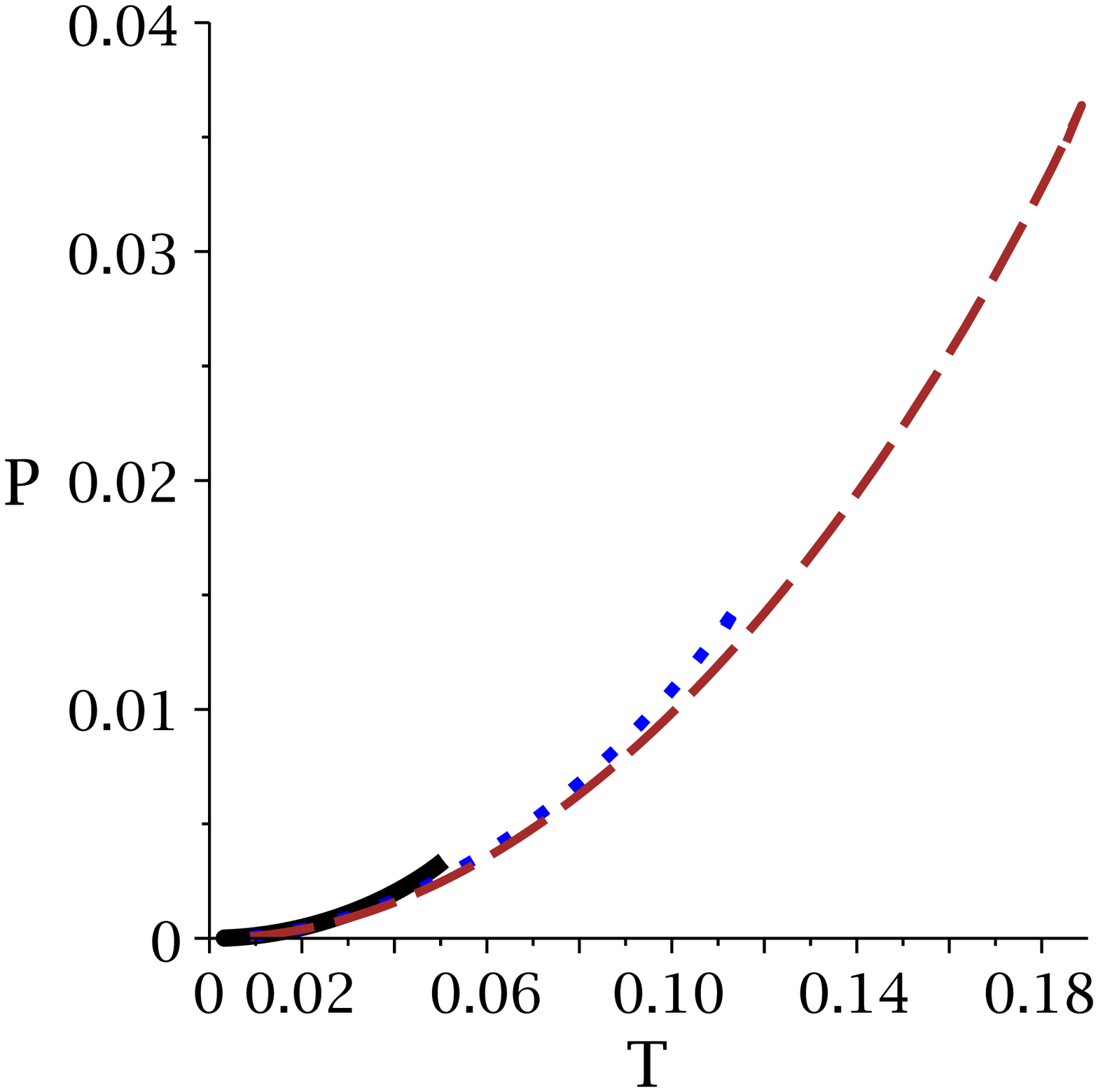}%
\end{array}
$%
\caption{Left panel: $P$ versus $r_{+}$ diagrams; Right panel: $P$ versus $T$
diagrams. \newline
for $k=1$, $l=1$ and $q=1.1$; $n=4$ (continues line), $n=5$ (dashed line)
and $n=6$ (dotted line). }
\label{Fig11}
\end{figure}

Evidently, critical pressure is a decreasing function of the electric charge
while its corresponding critical horizon radius is an increasing function of
this parameter (left panel of Fig. \ref{Fig6}). As for the effects of
dimensionality, one can see that both critical horizon radius and pressure
are increasing functions of this parameter (left panel of Fig. \ref{Fig7}).
The effects of nonlinearity parameter should be separated into two branches;
for $0.5<s<1,$ critical pressure is a decreasing function of $s,$ while the
corresponding critical horizon is an increasing function of it (left panel
of Fig. \ref{Fig8}). On the contrary, for $1<s$, critical pressure is an
increasing function of the nonlinearity parameter whereas its critical
horizon radius is a decreasing function of it (left panel of Fig. \ref{Fig9}%
). These opposite behaviors for these two branches, highlight the
differences between this nonlinear electromagnetic field and
Born-Infeld like in which the effects of nonlinearity parameter
are fixed.

As for CIM case, evidently, critical temperature and pressure are
decreasing functions of the electric charge, while the critical
horizon radius is an increasing function of it (Fig. \ref{Fig10}).
The effects of dimensionality is opposite of the variation of
electric charge. Meaning that, the critical pressure and
temperature are increasing functions of the dimensionality while
the critical horizon radius is a decreasing function of this
parameter (Fig. \ref{Fig11}).

\subsection{van der Waals properties}

In this subsection, we will investigate some van der Waals like properties
of the solutions in the extended phase space. The presence of pressure in
thermodynamical quantities and having equation of state at hand, one is able
to extract volume expansion coefficient, speed of sound and isothermal
compressibility coefficient of these black holes.

The volume expansion coefficient is calculated at constant pressure and it
represents changes in volume of the black holes which takes place due to
heat transferring. The volume expansion coefficients for two cases of this
paper are given by \cite{comp}
\begin{equation}
\alpha =\frac{1}{V}\left( \frac{\partial V}{\partial T}\right) _{P}=\left\{
\begin{array}{cc}
\begin{array}{c}
\,{\frac{4\pi n\left( n-1\right) \,{r}_{+}^{n+1}}{16\,\pi \,P{r}%
_{+}^{n+2}-k\left( n-1\right) \left( n-2\right) {r}_{+}^{n}+{q}^{n}{2}%
^{n/2}\left( n\left( n-2\right) +{1}\right) {r}_{+}^{2}},} \\
\\
\end{array}
&
\begin{array}{c}
s=\frac{n}{2} \\
\\
\end{array}
\\
\frac{4\pi \,n\,\left( n-1\right) {r}_{+}^{{\frac{2\,s\left( n-2\right) +1}{%
2\,s-1}}}}{16\pi P\,{r}_{+}^{\,{\frac{2s\left( n-1\right) }{2\,s-1}}%
}-k\left( n-1\right) \left( n-2\right) {r}_{+}^{{\frac{2\left( s\left(
n-3\right) +1\right) }{2\,s-1}}}+\frac{\left( {q}^{2}\left( n-1\right)
\left( n-2\,s\right) ^{2\,}\right) ^{s}\left( 2\,s\left( n-2\right)
+1\right) }{\left( n-2\right) ^{s}\left( 2\,s-1\right) ^{2\,s}}}, & otherwise%
\end{array}%
\right. .
\end{equation}

The isothermal compressibility coefficient is calculated at constant
temperature and it represents the corresponding effects of variation in
volume with respect to change in pressure. Here, one can obtain this
coefficient as \cite{comp}
\begin{equation}
\kappa _{T}=-\frac{1}{V}\left( \frac{\partial V}{\partial P}\right)
_{T}=\left\{
\begin{array}{cc}
\begin{array}{c}
{\frac{16\pi n{r}_{+}^{n+2}}{16\,\pi \,P{r}_{+}^{n+2}-k\left( n-1\right)
\left( n-2\right) {r}_{+}^{n}+{2}^{n/2}{q}^{n}\left( n\left( n-2\right) +{1}%
\right) {r}_{+}^{2}}}, \\
\\
\end{array}
&
\begin{array}{c}
s=\frac{n}{2} \\
\\
\end{array}
\\
\frac{16\pi n\,{r}_{+}^{{\frac{2s\left( n-1\right) }{2\,s-1}}}}{16\pi \,P{r}%
_{+}^{{\frac{2s\left( n-1\right) }{2\,s-1}}}-k\left( n-1\right) \left(
n-2\right) {r}_{+}^{{\frac{2\left( s\left( n-3\right) +1\right) }{2\,s-1}}}+%
\frac{\left( {q}^{2}\left( n-1\right) \left( n-2\,s\right) ^{2}\right)
^{s}\left( 2\,s\left( n-2\right) +1\right) }{\left( n-2\right) ^{s}\left(
2\,s-1\right) ^{2\,s}}}, & otherwise%
\end{array}%
\right.
\end{equation}%
where we have employed
\begin{equation}
\left( \frac{\partial V}{\partial P}\right) _{T}\left( \frac{\partial P}{%
\partial T}\right) _{V}\left( \frac{\partial T}{\partial V}\right) _{P}=-1.
\end{equation}

The speed of sound in black holes does not carries its usual meaning. Here,
while the area of black holes is fixed, the speed of sound indicates a
breathing mode for variation of the volume with pressure. For obtaining
speed of sound, one should first calculate the homogenous density which for
two cases here, can be written as \cite{dolan1,dolan2}
\begin{equation}
\rho =\frac{M}{V}=\left\{
\begin{array}{cc}
\begin{array}{c}
P+{\frac{kn\left( n-1\right) }{16\pi \,{r}_{+}^{2}}}-{\frac{{2}^{n/2\,}{q}%
^{n}n\,\left( n-1\right) \ln \left( \frac{r_{+}}{l}\right) }{16\pi \,{r}%
_{+}^{n}},} \\
\\
\end{array}
&
\begin{array}{c}
s=\frac{n}{2} \\
\\
\end{array}
\\
P+{\frac{n\left( n-1\right) \,k}{16\pi \,{r}_{+}^{2}}}+\frac{n\,\left( {q}%
^{2}\left( n-1\right) \left( n-2\,s\right) ^{2\,}\right) ^{s}\left(
2\,s-1\right) ^{2}}{16{\pi }\left( n-2\right) ^{s}\left( 2\,s-1\right)
^{2\,s}\left( n-2\,s\right) {r}_{+}^{{\frac{2\,s\left( n-1\right) +1}{2\,s-1}%
}}}, & otherwise%
\end{array}%
\right. .
\end{equation}

Now, by employing the method of Refs. \cite{dolan1,dolan2}, we obtain
\begin{equation}
c_{s}^{-2}=\frac{\partial \rho}{\partial P }=1+\rho \kappa ,
\end{equation}%
in which confirms that speed of sound for these black holes is in the valid
region of $0\leq c_{s}^{2}\leq 1$.

Obtained volume expansion and isothermal compressibility coefficients share
identical denominator. Therefore, their divergencies are matched. Here, we
focus only on the volume expansion coefficient. In order to have a better
picture regarding the meaning of divergencies and the sign of this quantity,
we have plotted its diagrams with the heat capacity and Ricci scalar of HPEM
metric in Figs. \ref{Fig1}-\ref{Fig5} (dashed-dotted lines).

First of all, as one can see, the divergencies of volume expansion
coefficient and heat capacity are matched. This shows that phase
transition points that are observed in the heat capacity could be
detected by studying the divergencies of volume expansion
coefficient as well. In other words, the divergencies of volume
expansion coefficient mark points in which black holes go under
second order phase transition. In the absence of divergency for
the volume expansion coefficient, this quantity is positive
valued. On the contrary, when this quantity acquires divergences,
its sign changes at the divergence points. By taking a closer look
at the phase diagrams, one can see that the regions in which black
holes are stable, the volume expansion coefficient is positive
valued. On the contrary, in the unstable regions (negative heat
capacity), the sign of volume expansion coefficient is negative.
Here, we see that the sign of heat capacity and the volume
expansion coefficient are identical in physical regions (positive)
while in the non-physical region (negative temperature), there is
a differences between the sign of these two quantities. It is
worthwhile to mention that for the negative volume expansion
coefficient, speed of sound will exceed the speed of light which
is not physically acceptable. This indicates that between two
divergencies, no physical black hole solutions exist.

Evidently, different approaches toward studying critical behavior of the
black holes yield consisting results regarding the number and places of
phase transition points. But, as one can see, each one of the phase diagrams
carries specific information regarding thermodynamical structure and
properties of the black holes which could not be acquired by other phase
diagrams. Therefore, for having a better and more complete picture regrading
thermodynamical behaviour/structure/properties of black holes, studying
different phase diagrams is necessary and crucial.

\section{Conclusion}

In this paper, we have investigated thermodynamical structure of Einstein
black holes in the presence of PMI nonlinear electromagnetic field through
new techniques.

It was shown that the behavior of temperature and stability
conditions for these black holes were functions of different
parameters describing the phase structure of these black.
Interestingly, it was shown that the nonlinearity of system and
the electric charge are two opposing factors in phase structure of
these black holes. This behavior precisely highlights the
differences between this theory of nonlinear electromagnetic field
and Born-Infeld like theories. In Born-Infeld theories, the
effects of both nonlinearity parameter and electric charge are
similar. Here, we see that in PMI theory, these effects are in
opposite of each other.

Next, geometrical thermodynamics was employed to investigate
thermodynamical behavior of these black holes. It was pointed out
that employing the Ruppeiner, Weinhold and Quevedo metrics will
lead to anomaly while using HPEM metric, one obtains consistent
results. The properties of HPEM metric around bound and phase
transition points were highlighted as well. It was shown that
(un)changing the sign of thermodynamical Ricci scalar around
divergence points helps one to distinguish bound point from phase
transition ones.

In addition, the proportionality between cosmological constant and
thermodynamical pressure in denominator of the heat capacity was
employed to extract a relation for thermodynamical pressure
(independent of equation of state). Using the new pressure
relation, it was possible to study the critical behavior of black
hole system. Besides, the effects of different parameters on the
critical pressure and horizon radius were investigated and it was
shown that positivity of critical pressure and horizon radius were
subjects to the choices of different parameters. In other words,
it was possible to eliminate the existence of second order phase
transition in phase space of these black holes by specific choices
of different parameter.

Finally, we have conducted a study regarding thermodynamical
properties of these black holes including volume expansion
coefficient, speed of sound and isothermal compressibility
coefficient. We have pointed out that divergencies of the volume
expansion coefficient were matched with phase transition points in
the heat capacity. This leads to a coincidence between
divergencies of the Ricci scalar of HPEM and volume expansion
coefficient. Furthermore, we showed that the sign of volume
expansion coefficient were same as that of the heat capacity in
physical regions ($T>0$) while in the non-physical region ($T<0$),
these two signs were opposite.

The thermodynamical behavior that we observed here highly depends
on choices of the nonlinearity parameter. In specific regions of
the nonlinearity parameter, we observed a series of the effects
which were modified in other regions for nonlinearity parameter.
This characteristic behavior is not observed in Born-Infeld like
family of the nonlinear electromagnetic fields. Meaning that the
PMI theory not only differs from Born-Infeld like theories in
form, but the characteristic behavior is different and only
observed in this theory of the nonlinear electromagnetic field.
These modifications in thermodynamical behavior of the black holes
signal the fact that evolution of the black holes in this theory
is different from other charged black holes. Therefore, the trace
of these differences may be observable in properties such as quasi
normal modes, gravitational waves and Hawking radiation and
provide interesting case studies in other aspects of physics such
as gauge/gravity duality.

\begin{acknowledgements}
We would like to thank the referees for the valuable comments. We
also wish to thank Shiraz University Research Council. This work
has been supported financially by the Research Institute for
Astronomy and Astrophysics of Maragha, Iran.
\end{acknowledgements}

\end{document}